\pdfoutput=1
\documentclass{article}
\usepackage{arxiv}
\usepackage[authoryear,round]{natbib}
\usepackage[utf8]{inputenc}
\usepackage[T1]{fontenc}
\usepackage{amsmath, amsfonts, amssymb, amsthm}
\usepackage{graphicx}
\usepackage{booktabs}
\usepackage{hyperref}
\usepackage{bm}
\usepackage{algorithm}
\usepackage{algpseudocode}
\usepackage{microtype}

\newtheorem{definition}{Definition}
\newtheorem{proposition}{Proposition}

\title{Parametric Modal SARIMA Models with Skewed Distributions}

\author{
  Christian E.\ Galarza \\
  Escuela Superior Polit\'ecnica del Litoral \\
  Guayaquil, Ecuador \\
  \texttt{chedgala@espol.edu.ec}
  \And
  Sergio Luis Mercado Londo\~{n}o\thanks{Corresponding author.} \\
  Universidad de Antioquia \\
  Medell\'{\i}n, Colombia \\
  \texttt{luis.mercado@udea.edu.co}
  \And
  V\'{\i}ctor H.\ Lachos \\
  University of Connecticut \\
  Storrs, CT, USA \\
  \texttt{hlachos@uconn.edu}
}

\begin{document}
\maketitle

\begin{abstract}
Seasonal autoregressive integrated moving-average (SARIMA) models typically target the conditional mean of a time series under a Gaussian error assumption. This approach may be inadequate when the innovations are asymmetric because the conditional mean no longer represents the most likely outcome.
We propose the \emph{Modal SARIMA}$(p,d,q)\times(P,D,Q)_{s}$ model, which links the conditional mode, rather than the conditional mean, to a multiplicative seasonal ARIMA structure. The innovations follow the Fernandez--Steel class of skewed distributions (SKD), whose location parameter is the mode, and the non-seasonal Modal ARIMA$(p,d,q)$ model is the special case without seasonal terms.
We give the conditional log-likelihood for the Skew-Normal, Skewed Student-$t$, and Skewed Laplace members, establish the stationarity and invertibility conditions on the product polynomials, show that the recursive modal forecast is the joint mode of the future trajectory, and characterize the correction that yields the marginal $h$-step conditional mode.
Monte Carlo experiments show that conditional maximum likelihood and its Wald inference are reliable when the seasonal factors are well separated, that the test of symmetry holds its level from about 200 observations, and that the modal estimator of the dynamic coefficients is as efficient as Gaussian least squares under symmetry and 1.2 to 2.6 times more efficient under asymmetry or heavy tails. We also show, analytically and by simulation, that nearly cancelling seasonal factors amplify the effect of the starting values, and that beyond one step the recursive forecast must be corrected to remain a forecast of the most probable value.
In an application to the Canadian lynx trappings, a seasonal component for the ten-year cycle improves the non-seasonal fit for every member of the SKD family.
The methods are implemented in the R package \texttt{ModalForecast}.
\end{abstract}


\section{Introduction}

Time series in economics, epidemiology, climatology, energy markets, and many other fields frequently exhibit two distinctive features: \emph{seasonality}, or recurring patterns at fixed calendar intervals, and \emph{distributional asymmetry}, or skewness in the conditional distribution of the observations.
Seasonal autoregressive integrated moving-average (SARIMA) models have become the standard tool for dealing with the former since the seminal work of \citet{BoxJenkins1976}, in which seasonal differencing and seasonal autoregressive (AR) and moving-average (MA) components are introduced to account for the periodic dynamics.
However, these models invariably target the conditional \emph{mean} of the series under a Gaussian error assumption.
When the error distribution is skewed, the conditional mean differs from the conditional mode---the value of highest probability density---and mean-based forecasts therefore no longer represent the single most likely outcome.

Modal regression, which models the conditional mode as a function of covariates, has been developed mainly for independent observations \citep{YaoLi2014, Xiang2026}.
In a parametric setting, \citet{GalarzaLachos2017} introduced the SKD (Skewed Distribution) family based on the Fernandez--Steel parameterization \citep{FS1998}. This unified class comprises the Skew-Normal, Skewed Student-$t$, and Skewed Laplace distributions, with a location parameter that coincides exactly with the distributional mode.
A time series model whose innovations belong to this family and have mode zero has a systematic component that is the conditional mode of the series, and therefore generates modal predictions directly.

This paper proposes such a model for seasonal time series.
The Modal SARIMA$(p,d,q)\times(P,D,Q)_{s}$ model, where $s$ denotes the seasonal period, links the conditional mode to a multiplicative seasonal ARIMA structure with SKD innovations, and it contains the non-seasonal Modal ARIMA$(p,d,q)$ model as the special case without seasonal terms.
Our contribution is threefold.
First, we formulate the model and derive the conditional-mode equation using multiplicative lag-operator polynomials that interleave regular and seasonal AR and MA terms.
Second, we derive the conditional log-likelihood for the three members of the SKD family, in which the regular and seasonal structure enters only through a recursion for the modal residuals, and we describe its maximization and the associated inference.
Third, we establish the stationarity and invertibility conditions governing the product polynomials and quantify the effect of the starting values of the recursion on the estimator.
In addition, we study the interpretation of the recursive modal forecasts, showing that they correspond to the joint mode of the future trajectory, and we characterize the correction needed to recover the marginal conditional mode at each horizon.
The proposed methodology is implemented in the R package \texttt{ModalForecast}, which provides practitioners with a convenient tool for applications in a range of fields.

The rest of the paper is organized as follows.
Section~\ref{sec:prelim} reviews the SKD family.
Section~\ref{sec:modal_sarima} presents the Modal SARIMA model.
Section~\ref{sec:likelihood} derives the log-likelihood under each member of the SKD family and discusses maximum likelihood (ML) estimation.
Section~\ref{sec:conditions} establishes the stationarity and invertibility conditions of the seasonal model.
In Section~\ref{sec:forecasting}, modal forecasting is discussed, with emphasis on the distinction between the joint modal trajectory and the marginal conditional mode.
Section~\ref{sec:simulation} reports Monte Carlo experiments on parameter recovery and inference, the comparison with Gaussian SARIMA, the starting transient, model selection, contamination, and forecasting.
Section~\ref{sec:application} presents the application to the Canadian lynx trappings.
Finally, Section~\ref{sec:conclusion} concludes with a discussion and suggests possible directions for future research.
The proofs are collected in Appendix~\ref{app:proofs}.

\section{The SKD Family}\label{sec:prelim}

The SKD family introduced by \citet{GalarzaLachos2017} unifies three distributions under the Fernandez--Steel reparameterization.
Let $y \sim \text{SKD}(\mu, \sigma, \gamma; g)$, where $\mu$ is the mode, $\sigma > 0$ is a scale parameter, $\gamma > 0$ is a skewness parameter, and $g(\cdot)$ denotes a symmetric, unimodal kernel density with its maximum at the origin.
Defining $p = 1/(\gamma^2 + 1)$, the probability density function (pdf) of $y$ is given by
\begin{equation}\label{eq:skd_general}
    f(y \mid \mu, \sigma, \gamma) = \frac{4\,p\,(1-p)}{\sigma}\,g\!\left(2\,\rho_p\!\left(\frac{y - \mu}{\sigma}\right)\right),
\end{equation}
where $\rho_p(u) = u\,(p - \mathbb{I}(u < 0))$ is the check (or quantile) function evaluated at $p$.
Note that, since $\int_{-\infty}^{\infty} g(2\rho_p(u))\,du = \{4p(1-p)\}^{-1}$ for any symmetric density $g$, the constant $4p(1-p)/\sigma$ normalizes~\eqref{eq:skd_general}.

The three members of the family are determined by the choice of kernel $g$; throughout, we write $z = (y - \mu)/\sigma$.
The \textbf{Skew-Normal} (SN) distribution corresponds to $g = \phi$, the standard normal density, and its pdf is
\begin{equation}\label{eq:skn}
    f(y \mid \mu, \sigma, \gamma) = \frac{4\,p\,(1-p)}{\sigma\sqrt{2\pi}}\exp\!\left(-2\,\rho_p^2(z)\right).
\end{equation}
It is important to note that this is the two-piece (Fernandez--Steel) normal distribution and not the Skew-Normal distribution of Azzalini.
Equivalently, it can be written in the Fernandez--Steel form
\begin{equation}\label{eq:skn_fs}
    f(y \mid \mu, \sigma, \gamma) = \frac{2}{\tilde{\sigma}(\gamma + \gamma^{-1})}
    \begin{cases}
        \phi\!\left(\dfrac{y - \mu}{\tilde{\sigma} \gamma}\right), & y \ge \mu, \\[0.8em]
        \phi\!\left(\dfrac{(y - \mu)\gamma}{\tilde{\sigma}}\right), & y < \mu,
    \end{cases}
    \qquad \tilde{\sigma} = \frac{\sigma\,(\gamma + \gamma^{-1})}{2},
\end{equation}
so that the scale parameter $\tilde{\sigma}$ of the Fernandez--Steel form differs from $\sigma$ by the factor $(\gamma + \gamma^{-1})/2$.

The \textbf{Skewed Student-$t$} (St) distribution corresponds to $g = t_\nu$, the Student-$t$ density with $\nu > 0$ degrees of freedom; $\nu > 2$ ensures a finite variance. Its pdf is
\begin{equation}\label{eq:skt}
    f(y \mid \mu, \sigma, \gamma, \nu) = \frac{4\,p\,(1-p)}{\sigma}\,\frac{\Gamma\!\left(\frac{\nu+1}{2}\right)}{\Gamma\!\left(\frac{\nu}{2}\right)\sqrt{\pi\nu}}\left(1 + \frac{4\,\rho_p^2(z)}{\nu}\right)^{-(\nu+1)/2}.
\end{equation}

Finally, the \textbf{Skewed Laplace} (SL) distribution corresponds to $g(u) = \tfrac{1}{2}e^{-|u|}$, the standard Laplace density, and has pdf
\begin{equation}\label{eq:skl}
    f(y \mid \mu, \sigma, \gamma) = \frac{2\,p\,(1-p)}{\sigma}\exp\!\left(-2\,\rho_p(z)\right).
\end{equation}

In all three cases, the kernel $g$ is unimodal and attains its maximum at the origin. Therefore, the density $f(y \mid \mu, \sigma, \gamma)$ attains its unique maximum at $y = \mu$, and $\mu$ is exactly the mode of the distribution.
Moreover, $P(y < \mu) = p$, so that the mode $\mu$ is also the $p$-th quantile of $y$; for $\gamma > 1$ the right tail is heavier and the mean exceeds the mode.

\section{The Modal SARIMA Model}\label{sec:modal_sarima}

Let $\{Y_t\}_{t=1}^{T}$ be the observed series and $w_t = (1 - B)^d\,(1 - B^s)^D\,Y_t$ the series after $d$ regular and $D$ seasonal differences, of length $T^{*} = T - d - Ds$, where $B$ is the backshift operator and $s$ the seasonal period.
In the usual SARIMA notation \citep{BoxJenkins1976}, $\phi(B) = 1 - \phi_1 B - \cdots - \phi_p B^p$ and $\theta(B) = 1 + \theta_1 B + \cdots + \theta_q B^q$ are the regular autoregressive and moving-average polynomials, and $\Phi(B^s) = 1 - \Phi_1 B^s - \cdots - \Phi_P B^{Ps}$ and $\Theta(B^s) = 1 + \Theta_1 B^s + \cdots + \Theta_Q B^{Qs}$ are their seasonal counterparts.

\begin{definition}[Modal SARIMA$(p,d,q)\times(P,D,Q)_s$]\label{def:modal_sarima}
The Modal SARIMA$(p,d,q)\times(P,D,Q)_s$ model specifies that
\begin{equation}\label{eq:modal_sarima}
    \phi(B)\,\Phi(B^s)\,w_t = c + \theta(B)\,\Theta(B^s)\,\epsilon_t,
\end{equation}
where $\epsilon_t \overset{\mathrm{iid}}{\sim} \mathrm{SKD}(0, \sigma, \gamma; g)$, the product polynomials satisfy the stationarity and invertibility conditions of Section~\ref{sec:conditions}, and $\mu_t = \mathrm{Mo}(w_t \mid \mathcal{F}_{t-1})$ denotes the conditional mode of $w_t$ given the filtration $\mathcal{F}_{t-1} = \sigma\{w_{t-1}, w_{t-2}, \ldots\}$.
\end{definition}

Causality and invertibility are both essential to this definition. Causality makes $\epsilon_t$ independent of $\mathcal{F}_{t-1}$, and invertibility makes each past innovation $\epsilon_{t-k}$ a convergent linear function of $w_{t-k}, w_{t-k-1}, \ldots$, so that $\mu_t$ is $\mathcal{F}_{t-1}$-measurable; together they give $w_t \mid \mathcal{F}_{t-1} \sim \mathrm{SKD}(\mu_t, \sigma, \gamma; g)$ and $\mu_t = \mathrm{Mo}(w_t \mid \mathcal{F}_{t-1})$.
Without invertibility the residual recursion of Section~\ref{subsec:recursion} is unstable: its errors grow geometrically when $\beta(z)$ has a zero inside the unit circle and do not die out when it has a zero on it.

Expanding the products,
\begin{equation}\label{eq:products}
    \alpha(B) = \phi(B)\,\Phi(B^s) = 1 - \sum_{k=1}^{p + Ps} \alpha_k B^k,
    \qquad
    \beta(B) = \theta(B)\,\Theta(B^s) = 1 + \sum_{k=1}^{q + Qs} \beta_k B^k,
\end{equation}
the model reads
\begin{equation}\label{eq:mode_explicit}
    w_t = c + \sum_{k=1}^{p+Ps} \alpha_k\,w_{t-k} + \epsilon_t + \sum_{k=1}^{q+Qs} \beta_k\,\epsilon_{t-k},
\end{equation}
so that the conditional mode is
\begin{equation}\label{eq:mu_t_sarima}
    \mu_t = c + \sum_{k=1}^{p+Ps} \alpha_k\,w_{t-k} + \sum_{k=1}^{q+Qs} \beta_k\,\epsilon_{t-k},
\end{equation}
with $\epsilon_t = w_t - \mu_t$ and $\mathrm{Mo}(\epsilon_t) = 0$.
The products create cross terms between regular and seasonal lags; in a Modal SARIMA$(1,1,1)\times(1,1,1)_{12}$ model, for instance,
\[
    \mu_t = c + \phi_1 w_{t-1} + \Phi_1 w_{t-12} - \phi_1\Phi_1 w_{t-13} + \theta_1 \epsilon_{t-1} + \Theta_1 \epsilon_{t-12} + \theta_1\Theta_1 \epsilon_{t-13}.
\]
Setting $P = D = Q = 0$ gives the non-seasonal special case, which we call the Modal ARIMA$(p,d,q)$ model,
\begin{equation}\label{eq:arima_lag}
    \phi(B)\,(1 - B)^d\,Y_t = c + \theta(B)\,\epsilon_t.
\end{equation}

\section{Log-Likelihood and Maximum Likelihood Estimation}\label{sec:likelihood}

The model is estimated by conditional maximum likelihood.
Because the innovations are independent, the log-likelihood is a sum of SKD log-densities evaluated at the modal residuals, and the regular and seasonal structure enters only through the recursion that defines these residuals.

\subsection{Modal Residual Recursion}\label{subsec:recursion}

Given the doubly differenced series $w_t$ and a set of initial values (discussed below), the modal residuals are computed through the recursion
\begin{equation}\label{eq:recursion}
    \epsilon_t = w_t - \mu_t = w_t - c - \sum_{k=1}^{p+Ps} \alpha_k\,w_{t-k} - \sum_{k=1}^{q+Qs} \beta_k\,\epsilon_{t-k}.
\end{equation}
Let the differenced series be indexed as $w_1, \ldots, w_{T^{*}}$ and let $t_0 = p + Ps$ denote the maximal autoregressive lag.
Since $w_t$ is unobserved for $t \le 0$, the recursion requires a starting rule, and we consider two options.
Under the \emph{conditional start}, we condition on the first $t_0$ observations, set $\epsilon_t = 0$ for $t \le t_0$, and evaluate~\eqref{eq:recursion} for $t = t_0+1, \ldots, T^{*}$, so that the likelihood uses the index set $\mathcal{T} = \{t_0+1, \ldots, T^{*}\}$ with $n = T^{*} - t_0$ terms.
Under the \emph{zero pre-sample} rule, we set $w_t = \epsilon_t = 0$ for $t \le 0$ and evaluate~\eqref{eq:recursion} for $t = 1, \ldots, T^{*}$, so that $\mathcal{T} = \{1, \ldots, T^{*}\}$ and $n = T^{*}$.
Both rules fix finitely many pre-sample values, and by Proposition~\ref{prop:invertibility}(c) the error they induce in the later residuals decays geometrically under invertibility; back-forecasting procedures analogous to those used in standard SARIMA estimation may also be used.
The transient does not alter the first-order asymptotic behaviour of the estimator, but it displaces it by a term of order $n^{-1}$ (Proposition~\ref{prop:start}), which is material when autoregressive and moving-average factors nearly cancel (Section~\ref{subsec:mc_transient}).

\subsection{Log-Likelihood and Its Maximization under the SKD Family}\label{subsec:loglik}

\begin{proposition}[Conditional log-likelihood]\label{prop:loglik}
Let
\[
    \bm{\Theta} = (c, \phi_1, \ldots, \phi_p, \theta_1, \ldots, \theta_q, \Phi_1, \ldots, \Phi_P, \Theta_1, \ldots, \Theta_Q, \sigma, \gamma)'
\]
denote the full parameter vector (augmented with $\nu$ for the Student-$t$ case).
For the index set $\mathcal{T}$ with $n = |\mathcal{T}|$ terms defined in Section~\ref{subsec:recursion}, the log-density of $(w_t)_{t \in \mathcal{T}}$ conditional on the pre-sample values fixed by the starting rule is
\begin{equation}\label{eq:loglik_general}
    \ell(\bm{\Theta}) = \sum_{t \in \mathcal{T}} \log f(\epsilon_t \mid 0, \sigma, \gamma; g),
\end{equation}
where $f(\cdot \mid 0, \sigma, \gamma; g)$ is the SKD density~\eqref{eq:skd_general} evaluated at zero mode, and $\epsilon_t$ is obtained from the recursion~\eqref{eq:recursion}.
The orders $(p,q)\times(P,Q)_s$ enter $\ell$ through $\epsilon_t$ and, under the conditional start, through $\mathcal{T}$, because $t_0 = p + Ps$; log-likelihoods and information criteria of different orders must therefore be computed on a common index set.
\end{proposition}

The proofs of this and the following propositions are given in Appendix~\ref{app:proofs}.
For completeness, we provide the explicit log-likelihood for each member of the family.

\paragraph{Skew-Normal.}
The log-likelihood is given by
\begin{equation}\label{eq:loglik_sn}
    \ell_{\mathrm{SN}}(\bm{\Theta}) = n\!\left[\log(4p(1-p)) - \log\sigma - \tfrac{1}{2}\log(2\pi)\right] - 2\sum_{t \in \mathcal{T}} \rho_p^2\!\left(\frac{\epsilon_t}{\sigma}\right).
\end{equation}
In the Fernandez--Steel parameterization~\eqref{eq:skn_fs}, the same log-likelihood reads
\begin{equation*}
    \ell_{\mathrm{SN}} = n\!\left[\log 2 - \log\tilde{\sigma} - \log(\gamma + \gamma^{-1}) - \tfrac{1}{2}\log(2\pi)\right]
    - \frac{1}{2}\sum_{\substack{t \in \mathcal{T}\\ \epsilon_t \ge 0}} \left(\frac{\epsilon_t}{\tilde{\sigma}\gamma}\right)^{\!2}
    - \frac{1}{2}\sum_{\substack{t \in \mathcal{T}\\ \epsilon_t < 0}} \left(\frac{\epsilon_t\,\gamma}{\tilde{\sigma}}\right)^{\!2}.
\end{equation*}
Note that the constant $-\tfrac{n}{2}\log(2\pi)$ does not affect the maximizer, but it must be retained when comparing the SN fit with the Student-$t$ and Laplace fits through the log-likelihood or through information criteria such as the Akaike information criterion (AIC) and the Bayesian information criterion (BIC).

\paragraph{Skewed Student-$t$.}
The log-likelihood is given by
\begin{equation}\label{eq:loglik_st}
\begin{split}
    \ell_{\mathrm{St}}(\bm{\Theta}) &= n\!\left[\log\Gamma\!\left(\tfrac{\nu+1}{2}\right) - \log\Gamma\!\left(\tfrac{\nu}{2}\right) - \tfrac{1}{2}\log(\pi\nu) + \log(4p(1-p)) - \log\sigma\right] \\
    &\quad - \frac{\nu+1}{2}\sum_{t \in \mathcal{T}} \log\!\left(1 + \frac{4\,\rho_p^2(\epsilon_t/\sigma)}{\nu}\right),
\end{split}
\end{equation}
where $p = 1/(\gamma^2 + 1)$ and $\rho_p(u) = u\,(p - \mathbb{I}(u < 0))$.

\paragraph{Skewed Laplace.}
The log-likelihood is given by
\begin{equation}\label{eq:loglik_sl}
    \ell_{\mathrm{SL}}(\bm{\Theta}) = n\!\left[\log(2p(1-p)) - \log\sigma\right] - 2\sum_{t \in \mathcal{T}} \rho_p\!\left(\frac{\epsilon_t}{\sigma}\right).
\end{equation}

In all three cases, the regular and seasonal coefficients enter only through the residuals $\epsilon_t$ of the recursion~\eqref{eq:recursion}.
Moreover, because all three members share the parameterization~\eqref{eq:skd_general}, $\gamma$ has the same meaning across the three fits, since $P(\epsilon_t < 0) = 1/(1 + \gamma^2)$ for every kernel, whereas the dispersion implied by $\sigma$ depends on $g$: for $\gamma = 1$ the standard deviation of $\epsilon_t$ is $\sigma$, $\sqrt{2}\,\sigma$ and $\sigma\sqrt{\nu/(\nu - 2)}$ for the SN, SL and St members.

ML estimation is carried out by minimizing $-\ell(\bm{\Theta})$.
We apply logarithmic transformations to $\sigma$ and $\gamma$ (and to $\nu - 2$ in the Student-$t$ case) to enforce the positivity constraints. The parameter vector for unconstrained optimization then becomes
\begin{equation}
    \bm{\vartheta} = (c, \phi_1, \ldots, \phi_p, \theta_1, \ldots, \theta_q, \Phi_1, \ldots, \Phi_P, \Theta_1, \ldots, \Theta_Q, \log\sigma, \log\gamma)'.
\end{equation}
Initial values are obtained from the standard (mean-based) SARIMA fit via conditional sum of squares (CSS), which provides $\hat{\phi}_i^{(0)}, \hat{\theta}_j^{(0)}, \hat{\Phi}_I^{(0)}, \hat{\Theta}_J^{(0)}, \hat{\sigma}^{(0)}$, together with the symmetric initialization $\gamma^{(0)} = 1$ (that is, $\log\gamma^{(0)} = 0$) and, for the Student-$t$ distribution, $\nu^{(0)} = 5$ (that is, $\log(\nu^{(0)} - 2) = \log 3$).
The BFGS quasi-Newton algorithm is then applied with analytical gradients, obtained by the chain rule from the derivatives of the residual recursion with respect to the expanded polynomial coefficients.
The complete estimation procedure is summarized in Algorithm~\ref{alg:mle}.

\begin{algorithm}[!htb]
\caption{ML estimation of the Modal SARIMA$(p,d,q)\times(P,D,Q)_s$ model}
\label{alg:mle}
\begin{algorithmic}[1]
\Require Series $\{Y_t\}_{t=1}^{T}$, orders $(p,d,q)\times(P,D,Q)_s$, kernel $g$ (SN, St or SL), tolerance $\varepsilon$
\State Compute the doubly differenced series $w_t = (1-B)^d(1-B^s)^D Y_t$, $t = 1, \ldots, T^{*}$
\State \textbf{Initialize:} fit a Gaussian SARIMA$(p,d,q)\times(P,D,Q)_s$ by CSS to obtain $\hat{\phi}_i^{(0)}$, $\hat{\theta}_j^{(0)}$, $\hat{\Phi}_I^{(0)}$, $\hat{\Theta}_J^{(0)}$, $\hat{\sigma}^{(0)}$; set $\gamma^{(0)} = 1$ and, for the St kernel, $\nu^{(0)} = 5$
\State Set $\bm{\vartheta}^{(0)}$ from the initial values, with $\log\sigma^{(0)}$, $\log\gamma^{(0)} = 0$ and, for the St kernel, $\log(\nu^{(0)} - 2) = \log 3$
\Repeat
    \State Expand $\alpha(B) = \phi(B)\,\Phi(B^s)$ and $\beta(B) = \theta(B)\,\Theta(B^s)$ at $\bm{\vartheta}^{(k)}$ to obtain $\alpha_1, \ldots, \alpha_{p+Ps}$ and $\beta_1, \ldots, \beta_{q+Qs}$
    \State Compute the modal residuals $\epsilon_t$, $t \in \mathcal{T}$, through the recursion~\eqref{eq:recursion} with the chosen starting rule
    \State Evaluate $\ell(\bm{\vartheta}^{(k)})$ from~\eqref{eq:loglik_sn}, \eqref{eq:loglik_st} or~\eqref{eq:loglik_sl}, and its analytical gradient $\nabla\ell(\bm{\vartheta}^{(k)})$
    \State Check the roots of $\alpha(z)$ and $\beta(z)$ (Section~\ref{sec:conditions}) and penalize $-\ell$ if any root approaches the unit circle
    \State Update $\bm{\vartheta}^{(k+1)}$ by a BFGS quasi-Newton step on $-\ell$
\Until{$|\ell(\bm{\vartheta}^{(k+1)}) - \ell(\bm{\vartheta}^{(k)})| < \varepsilon$ or the maximum number of iterations is reached}
\If{$g$ is the Laplace kernel}
    \State Refine the solution by a derivative-free (Nelder--Mead) search started at the BFGS solution
\EndIf
\State Back-transform: $\hat{\sigma} = \exp(\log\hat{\sigma})$, $\hat{\gamma} = \exp(\log\hat{\gamma})$ and, for the St kernel, $\hat{\nu} = 2 + \exp(\log(\hat{\nu} - 2))$
\Ensure $\hat{\bm{\Theta}}$, $\ell(\hat{\bm{\Theta}})$ and the observed information at $\hat{\bm{\Theta}}$ for the standard errors
\end{algorithmic}
\end{algorithm}

Several features of Algorithm~\ref{alg:mle} warrant further discussion.
For the Skew-Normal and Skewed Student-$t$ distributions, the log-likelihood is continuously differentiable because $\rho_p^2(u)$ is $C^1$ at $u = 0$.
For the Skewed Laplace distribution, however, $\rho_p(u)$ has a kink at $u = 0$, so that $\ell_{\mathrm{SL}}$ is not differentiable whenever some $\epsilon_t = 0$, and gradient-based methods may stall at non-optimal points; this is the reason for the derivative-free refinement started at the BFGS solution in the final step of the algorithm.
In all cases, the CSS initialization is important, since seasonal coefficients initialized at zero may lead the optimizer to a local maximum.
Standard errors are obtained from the inverse of the observed information at $\hat{\bm{\Theta}}$, that is, the Hessian of $-\ell$; because $\ell_{\mathrm{SL}}$ is not twice differentiable everywhere, its observed information is instead estimated by the outer product of the per-observation score vectors, and the same estimator is used whenever the Hessian fails to be positive definite.

\section{Stationarity and Invertibility Conditions}\label{sec:conditions}

This section gives the conditions on the product polynomials under which the model is stationary and invertible.
Throughout, $\alpha(z) = \phi(z)\,\Phi(z^s)$ and $\beta(z) = \theta(z)\,\Theta(z^s)$ denote the product polynomials~\eqref{eq:products}, evaluated at a complex argument $z$, and we assume that
\begin{enumerate}
    \item[(A1)] $\alpha(z)$ and $\beta(z)$ have no common zeros, and
    \item[(A2)] $\mathrm{E}(\epsilon_t^2) < \infty$.
\end{enumerate}
Assumption (A2) holds for the Skew-Normal and Skewed Laplace members, and for the Skewed Student-$t$ member when $\nu > 2$.
Note that the innovations have mode zero but, when $\gamma \neq 1$, a nonzero mean $m = \mathrm{E}(\epsilon_t)$; this does not affect the conditions below.

The roots of $\alpha(z)$ are the union of the roots of $\phi(z)$ and of $\Phi(z^s)$, whether or not these factors share roots.
Moreover, each root $\zeta_0$ of $\Phi(\zeta)$, viewed as a polynomial in $\zeta$, generates the $s$ roots of $z^s = \zeta_0$, all of modulus $|\zeta_0|^{1/s}$, and $|\zeta_0|^{1/s} > 1$ if and only if $|\zeta_0| > 1$.
Hence
\begin{equation}\label{eq:root_conditions}
    \alpha(z) \neq 0 \ \text{for all } |z| \le 1
    \quad\Longleftrightarrow\quad
    \phi(z) \neq 0 \ \text{and} \ \Phi(z) \neq 0 \ \text{for all } |z| \le 1,
\end{equation}
and the same equivalence holds for $\beta(z)$ with $\theta(z)$ and $\Theta(\zeta)$.

\begin{proposition}[Causal stationarity]\label{prop:stationarity}
Under (A1) and (A2), the model~\eqref{eq:modal_sarima} has a stationary solution $\{w_t\}$ that is causal, that is,
\begin{equation}\label{eq:causal}
    w_t = \kappa + \sum_{j=0}^{\infty} \psi_j\,(\epsilon_{t-j} - m), \qquad \sum_{j=0}^{\infty} |\psi_j| < \infty,
\end{equation}
for a constant $\kappa$, if and only if $\alpha(z) \neq 0$ for all $|z| \le 1$, that is, if and only if all roots of $\phi(z)$ and of $\Phi(\zeta)$ lie outside the unit circle.
In that case the solution is unique, $\psi(z) = \sum_{j \ge 0} \psi_j z^j = \beta(z)/\alpha(z)$ for $|z| \le 1$, and $\kappa = \mathrm{E}(w_t) = \{c + \beta(1)\,m\}/\alpha(1)$.
\end{proposition}

\begin{proposition}[Invertibility]\label{prop:invertibility}
Let $\{w_t\}$ be the causal stationary solution of Proposition~\ref{prop:stationarity}.
Under (A1) and (A2), the innovations admit the representation
\begin{equation}\label{eq:invertible}
    \epsilon_t = \pi^{*} + \sum_{j=0}^{\infty} \pi_j\,w_{t-j}, \qquad \sum_{j=0}^{\infty} |\pi_j| < \infty,
\end{equation}
for a constant $\pi^{*}$, if and only if $\beta(z) \neq 0$ for all $|z| \le 1$, that is, if and only if all roots of $\theta(z)$ and of $\Theta(\zeta)$ lie outside the unit circle.
In that case:
\begin{enumerate}
    \item[(a)] $\pi(z) = \sum_{j \ge 0} \pi_j z^j = \alpha(z)/\beta(z)$ and $\pi^{*} = m - \pi(1)\,\kappa = -c/\beta(1)$;
    \item[(b)] the conditional mode $\mu_t$ in~\eqref{eq:mu_t_sarima} is a function of $w_{t-1}, w_{t-2}, \ldots$ only, and is therefore $\mathcal{F}_{t-1}$-measurable, so that $\mu_t = \mathrm{Mo}(w_t \mid \mathcal{F}_{t-1})$;
    \item[(c)] if the recursion~\eqref{eq:recursion} is started from arbitrary values, the difference $d_t = \hat\epsilon_t - \epsilon_t$ between the computed and the true residuals satisfies $|d_t| \le C\,\rho^{t}$ for any $\rho$ with $1/r < \rho < 1$, where $r > 1$ is the smallest modulus of the roots of $\beta(z)$ and $C$ is a random constant determined by the starting values and the pre-sample observations and innovations.
\end{enumerate}
\end{proposition}

Part~(c) justifies the statement in Section~\ref{subsec:recursion} that the choice of starting values affects the residuals only through a geometrically decaying transient. The bound holds uniformly over compact parameter sets on which the zeros of $\beta(z)$ have modulus at least some $r_0 > 1$, with $\rho \in (1/r_0, 1)$, so the transient does not affect the asymptotic behaviour of the conditional likelihood.

During numerical optimization, these constraints are enforced either by penalization, that is, by adding a large value to the negative log-likelihood whenever a root approaches the unit circle, or by reparameterization of the AR and MA coefficients into partial autocorrelation or reflection coefficient space.

The transient is nevertheless not innocuous in finite samples.
Let $\hat\epsilon_t(\bm{\Theta};a)$ denote the residuals computed from~\eqref{eq:recursion} under the conditional start with $\epsilon_t = a$ for $t \le t_0$, so that $a = 0$ in Section~\ref{subsec:recursion}, and let $\epsilon_t(\bm{\Theta})$ denote the residuals defined from the infinite past by~\eqref{eq:invertible}.
Write $\hat\ell_n(\bm{\Theta};a) = \sum_{t \in \mathcal{T}} \log f(\hat\epsilon_t(\bm{\Theta};a) \mid 0, \sigma, \gamma)$ and $\ell_n(\bm{\Theta}) = \sum_{t \in \mathcal{T}} \log f(\epsilon_t(\bm{\Theta}) \mid 0, \sigma, \gamma)$, and let $\bm{\eta}$ collect the distribution parameters $(\sigma, \gamma)$, augmented with $\nu$ for the Student-$t$ member.

\begin{proposition}[Starting value and the score]\label{prop:start}
Let $\bm{\Theta}_0$ be the true parameter and let (A1), (A2), and the root conditions of Propositions~\ref{prop:stationarity} and~\ref{prop:invertibility} hold.
Suppose that $s(e) = \partial \log f(e \mid 0, \bm{\eta}_0)/\partial e$ is Lipschitz and that $\bm{\lambda}(e) = \partial \log f(e \mid 0, \bm{\eta})/\partial \bm{\eta}\,|_{\bm{\eta}_0}$ satisfies $\|\bm{\lambda}(e + d) - \bm{\lambda}(e)\| \le L\,(1 + |e| + |d|)\,|d|$ for some constant $L$ and all $e$ and $d$.
Let $d_t = \hat\epsilon_t(\bm{\Theta}_0;a) - \epsilon_t$. Then:
\begin{enumerate}
    \item[(a)] $\mathrm{E}(d_t) = (a - m)\,H_t$ for $t \in \mathcal{T}$, where $H_t$ solves $\beta(B)\,H_t = 0$ for $t > t_0$ with $H_t = 1$ for $t \le t_0$; in particular, $\mathrm{E}(d_t) = 0$ for every $t$ when $a = m = \mathrm{E}(\epsilon_t)$;
    \item[(b)] $\Delta_n(a) = \nabla\hat\ell_n(\bm{\Theta}_0;a) - \nabla\ell_n(\bm{\Theta}_0)$ converges almost surely and in $L^1$, as $n \to \infty$, to a random vector $\Delta(a)$ with finite mean $b(a) = \mathrm{E}\{\Delta(a)\}$, and $\mathrm{E}\{\nabla\hat\ell_n(\bm{\Theta}_0;a)\} = b(a) + O(\rho^{\,n})$ for any $\rho$ as in Proposition~\ref{prop:invertibility}(c).
\end{enumerate}
\end{proposition}

The conditions on $s$ and $\bm{\lambda}$ hold for the Skew-Normal member, for which $s$ is piecewise linear and $\bm{\lambda}$ piecewise quadratic, and for the Skewed Student-$t$ member; they fail for the Skewed Laplace member, whose $s$ is discontinuous at zero.
Proposition~\ref{prop:start} has two practical consequences.
First, by the usual expansion $\hat{\bm{\Theta}} - \bm{\Theta}_0 \approx \{n\,I(\bm{\Theta}_0)\}^{-1}\,\nabla\hat\ell_n(\bm{\Theta}_0;a)$, where $I$ is the information per observation, the starting rule displaces the estimator by $\{n\,I(\bm{\Theta}_0)\}^{-1} \Delta_n(a)$, a term of order $n^{-1}$ with mean $\{n\,I(\bm{\Theta}_0)\}^{-1}\,b(a)$.
This is negligible relative to the sampling error of order $n^{-1/2}$, in agreement with Proposition~\ref{prop:invertibility}(c), unless $I(\bm{\Theta}_0)$ is nearly singular.
That is the case when autoregressive and moving-average factors nearly cancel: for a Gaussian ARMA(1,1) pair with factors $(1 - \Phi z)$ and $(1 + \Theta z)$, the per-observation information matrix of $(\Phi, \Theta)$, which does not depend on the innovation variance, has determinant $(\Phi + \Theta)^2/\{(1 - \Phi^2)(1 - \Theta^2)(1 + \Phi\Theta)^2\}$ \citep[whose moving-average coefficient is $-\Theta$]{BoxJenkins1976}, which vanishes as $\Theta \to -\Phi$.
Second, by~(a), $d_t - (a - m)\,H_t$ does not depend on $a$, so that $b(a)$ depends on $a$ only through $a - m$.
Under the conditional start $a = 0$, and
\[
    a - m = -m = -\,\sigma\,\mathrm{E}|X|\,\frac{\gamma^4 - 1}{2\gamma^2}, \qquad X \sim g,
\]
which is zero only when $\gamma = 1$.
Starting at the mode therefore adds to the displacement terms in $m$ that grow with the asymmetry of the innovations.
A start at $a = m$ removes these terms, but not a second-order contribution of the variance of the pre-sample innovations, which vanishes under symmetry but not otherwise (Appendix~\ref{app:proofs}); moreover, in practice $m = m(\sigma, \gamma)$ is evaluated at the current values of $\sigma$ and $\gamma$, so that $a$ varies with the parameters, a case not covered by Proposition~\ref{prop:start}.
Section~\ref{subsec:mc_transient} quantifies these effects.

\section{Modal Forecasting}\label{sec:forecasting}

At forecast origin $T$, the modal forecast path $\hat{\mu}_{T+1|T}, \ldots, \hat{\mu}_{T+H|T}$ is computed by iterating the recursion
\begin{equation}\label{eq:forecast}
    \hat{\mu}_{T+h|T} = c + \sum_{k=1}^{p+Ps} \alpha_k\,\tilde{w}_{T+h-k} + \sum_{k=1}^{q+Qs} \beta_k\,\tilde{\epsilon}_{T+h-k},
\end{equation}
where
\begin{equation}
    \tilde{w}_{T+h-k} =
    \begin{cases}
        w_{T+h-k}, & \text{if } T+h-k \le T, \\
        \hat{\mu}_{T+h-k|T}, & \text{if } T+h-k > T,
    \end{cases}
\end{equation}
and
\begin{equation}
    \tilde{\epsilon}_{T+h-k} =
    \begin{cases}
        \epsilon_{T+h-k}, & \text{if } T+h-k \le T, \\
        0, & \text{if } T+h-k > T,
    \end{cases}
\end{equation}
since the mode of $\epsilon_t$ is zero.
The seasonal lags extend the memory of the recursion to multiples of $s$.

The interpretation of~\eqref{eq:forecast} requires care, however, because the mode, unlike the mean, is not a linear operator.

\begin{proposition}[Joint modal trajectory]\label{prop:joint_mode}
For any horizon $H \ge 1$, the path $(\hat{\mu}_{T+1|T},\allowbreak \ldots,\allowbreak \hat{\mu}_{T+H|T})$ obtained from~\eqref{eq:forecast} is the joint mode of the conditional distribution of $(w_{T+1},\allowbreak \ldots,\allowbreak w_{T+H})$ given $\mathcal{F}_T$. In particular, $\hat{\mu}_{T+1|T} = \mathrm{Mo}(w_{T+1} \mid \mathcal{F}_T)$.
\end{proposition}

For $h \ge 2$, however, $\hat{\mu}_{T+h|T}$ is in general \emph{not} the marginal conditional mode of $w_{T+h}$.
Writing $\psi(B) = \beta(B)/\alpha(B) = \sum_{j \ge 0} \psi_j B^j$ with $\psi_0 = 1$, we have
\begin{equation}\label{eq:forecast_error}
    w_{T+h} - \hat{\mu}_{T+h|T} = \sum_{j=0}^{h-1} \psi_j\,\epsilon_{T+h-j},
\end{equation}
and the mode of a weighted sum of independent skewed variables with zero mode is generally different from zero.
Consequently, the marginal $h$-step modal forecast is given by
\begin{equation}\label{eq:marginal_mode}
    \mathrm{Mo}(w_{T+h} \mid \mathcal{F}_T) = \hat{\mu}_{T+h|T} + \delta_h,
    \qquad \delta_h = \mathrm{Mo}\Big(\sum_{j=0}^{h-1} \psi_j\,\epsilon_{T+h-j}\Big),
\end{equation}
where the density of the sum is the convolution of the scaled SKD densities and can be evaluated numerically (for example, by the fast Fourier transform) or by simulation.

\begin{proposition}[Marginal-mode correction]\label{prop:delta}
Let $\delta_h$ be defined by~\eqref{eq:marginal_mode}. Then:
\begin{enumerate}
    \item[(a)] For the SN and SL members, the density of $\sum_{j=0}^{h-1}\psi_j\,\epsilon_{T+h-j}$ is unimodal with a unique mode, so that $\delta_h$ is well defined.
    \item[(b)] $\delta_1 = 0$, and $\delta_h = 0$ for every $h$ when $\gamma = 1$.
    \item[(c)] If $\psi_1 < 0$, then $\delta_2 = 0$ for every $\gamma$ under the SN and SL members.
    \item[(d)] For the stationary series $w_t$, the density of the sum converges uniformly as $h \to \infty$, so that $\delta_h$ converges to the mode of $\sum_{j \ge 0}\psi_j\,\epsilon_j$ whenever this mode is unique.
\end{enumerate}
\end{proposition}

The Student-$t$ density is not log-concave, and for the St member $\delta_h$ is the global maximizer of the density of the sum; this density was unimodal in every case we computed, namely both designs of Section~\ref{sec:simulation} with $h \le 36$, $\nu \in \{3, 4, 10, 30\}$, and $\gamma \in \{0.4, 0.67, 1.5, 2.5\}$.
By~(c), $\delta_h$ depends on the signs of the weights $\psi_j$ as well as on $\gamma$: in the airline model $w_t = (1 + \theta_1 B)(1 + \Theta_1 B^s)\,\epsilon_t$ with $\theta_1 < 0$, for instance, $\delta_h = 0$ for $2 \le h \le s$.
In seasonal models the convergence in~(d) is stepwise: in the separated design of Section~\ref{subsec:mc_forecast}, $\delta_h$ changes by at most 3.3\% between $h = 6$ and $h = 12$, but by 46\% to 56\% from $h = 12$ to $h = 13$, when $\psi_{12} \approx 1$ enters, and its limit is about four to five times $\delta_{12}$.

The same reasoning applies to forecasts of the original series.
Undoing the differencing gives the joint modal trajectory of $(Y_{T+1}, \ldots, Y_{T+H})$, but the marginal mode of $Y_{T+h}$ must be obtained from the weights of $\psi^{*}(B) = \beta(B)/\{\alpha(B)(1-B)^d(1-B^s)^D\}$.
Because these weights do not decay when $d + D > 0$, the gap between the joint modal trajectory and the marginal mode of $Y_{T+h}$ can become substantial at long horizons.
Practitioners should therefore state which of the two targets is being forecast: the most probable \emph{trajectory}~\eqref{eq:forecast}, or the most probable \emph{value} at each horizon~\eqref{eq:marginal_mode}.
When the forecast is meant to be the most probable value at a horizon $h \ge 2$, the marginal mode~\eqref{eq:marginal_mode} is the appropriate point forecast: the simulation of Section~\ref{subsec:mc_forecast} shows that the uncorrected recursion can then be a poorer forecast of that value than the conditional mean.

Prediction intervals can be constructed in two ways.
Asymptotic intervals can be obtained by applying the delta method to the cumulative impulse-response weights of the expanded MA representation, with the quantiles of the forecast error~\eqref{eq:forecast_error} computed from the convolution of the fitted SKD densities.
Alternatively, bootstrap intervals can be obtained by generating $N$ forward simulation paths, drawing innovations from the estimated $\mathrm{SKD}(0, \hat{\sigma}, \hat{\gamma}; g)$, iterating the full seasonal recursion~\eqref{eq:mode_explicit}, and computing the empirical quantiles. The same simulated paths also provide an estimate of the marginal modal forecast~\eqref{eq:marginal_mode}.

\section{Monte Carlo Study}\label{sec:simulation}

\subsection{Design}\label{subsec:mc_design}

Six experiments assess the conditional ML estimator and the modal forecasts: (A) parameter recovery, Wald inference, and comparison with Gaussian SARIMA; (B) the effect of the starting transient; (C) the choice of the innovation distribution; (D) the choice of orders; (E) contamination; and (F) forecasting. All of them use the Modal SARIMA$(1,1,1)\times(1,1,1)_{12}$ model
\[
 (1-\phi B)(1-\Phi B^{12})\,w_t = c+(1+\theta B)(1+\Theta B^{12})\,\epsilon_t,
 \qquad w_t=(1-B)(1-B^{12})Y_t,
\]
with $c=0.20$, $\phi=0.50$, $\theta=0.30$, $\Phi=0.60$, and $\sigma=1$ in~\eqref{eq:skd_general}. Two values of the seasonal moving-average coefficient are considered. With $\Theta=0.40$ the seasonal factors $(1-0.60B^{12})$ and $(1+0.40B^{12})$ are well separated; with $\Theta=-0.40$ the factors $(1-0.60B^{12})$ and $(1-0.40B^{12})$ nearly cancel. We refer to these as the separated and the nearly cancelling designs. Experiment~A crosses both designs with eight innovation scenarios: Skew-Normal (SN) with $\gamma\in\{1,1.05,1.5,2.5\}$, Skewed Student-$t$ (ST) with $\nu=4$ and $\gamma\in\{1,1.05,1.5\}$, and Skewed Laplace (SL) with $\gamma=1.5$. The sample sizes are $n\in\{50,100,200,500,1000\}$, where $n$ is the length of the differenced series $w_t$; the undifferenced series has $n+13$ observations and the conditional likelihood uses $n-13$ residuals. Each cell has 1,000 replications in Experiment~A and 200 to 500 in the others, and 500 burn-in observations are discarded.

Estimation uses the conditional start of Section~\ref{subsec:recursion}. Gaussian conditional least squares is run from a neutral grid of nine starting values of $(\Phi,\Theta)$, none of which is the true value, and conditional ML is then run on the unconstrained scale from three starting values of $\log\gamma$, with the numerical bounds $|\log\gamma|\le 3$ and $\nu\le 202$. The SN and ST members use a quasi-Newton method and standard errors from the observed information and the delta method; the SL member adds a Nelder--Mead refinement and uses the outer product of scores. The experiments use an independent Python implementation of the conditional likelihood with numerical derivatives rather than the \texttt{ModalForecast} package.

No replication is discarded. A fit is called a boundary fit when an autoregressive or moving-average coefficient exceeds 0.98 in absolute value, $\hat\gamma\notin[0.25,4]$, or $\hat\nu>150$. The intercept and the coefficients are summarized on the natural scale. For $\gamma$, $\sigma$, and $\nu$, moments on the natural scale are determined by the numerical bounds whenever a fit reaches them, so we report the proportion of boundary fits and the median over all replications, and the bias, RMSE, and coverage for $\log\gamma$, $\log\sigma$, and $\log(\nu-2)$ over the non-boundary fits. With 1,000 replications, the Monte Carlo standard error of a coverage near 0.95 is 0.007. The optimizer converged in every replication of Experiment~A, usable standard errors were obtained in 99.9\% of them, and the maximized log-likelihood fell below its value at the true parameter in at most 0.6\% of the replications of any cell with $n\ge 100$. Complete results for all experiments are given in the supplement.

\subsection{Parameter recovery and Wald inference}\label{subsec:mc_recovery}

\begin{table}[!htbp]
\centering\small
\caption{Monte Carlo recovery of the intercept and of the autoregressive and seasonal coefficients, design with separated seasonal factors.}
\label{tab:mc_recovery}
\ifdefined\mctabbox\else\newsavebox\mctabbox\fi
\sbox\mctabbox{%
\begin{tabular}{ll *{12}{r}}
\toprule
& & \multicolumn{3}{c}{$c$} & \multicolumn{3}{c}{$\phi$} & \multicolumn{3}{c}{$\Phi$} & \multicolumn{3}{c}{$\Theta$}\\
\cmidrule(lr){3-5}\cmidrule(lr){6-8}\cmidrule(lr){9-11}\cmidrule(lr){12-14}
Scenario & $n$ & Bias & RMSE & Cov. & Bias & RMSE & Cov. & Bias & RMSE & Cov. & Bias & RMSE & Cov.\\
\midrule
SN, $\gamma=1$ & 100 & 0.025 & 0.642 & 0.885 & -0.034 & 0.153 & 0.910 & 0.023 & 0.106 & 0.902 & -0.043 & 0.163 & 0.905 \\
 & 500 & -0.007 & 0.224 & 0.939 & -0.007 & 0.057 & 0.952 & 0.000 & 0.044 & 0.955 & -0.004 & 0.054 & 0.951 \\
 & 1000 & -0.014 & 0.153 & 0.949 & -0.004 & 0.041 & 0.935 & 0.002 & 0.031 & 0.947 & -0.005 & 0.037 & 0.957 \\
\cmidrule{1-14}
SN, $\gamma=1.5$ & 100 & 0.082 & 1.075 & 0.832 & -0.026 & 0.150 & 0.853 & 0.029 & 0.105 & 0.825 & -0.056 & 0.161 & 0.826 \\
 & 500 & 0.025 & 0.298 & 0.951 & -0.004 & 0.053 & 0.939 & 0.007 & 0.039 & 0.947 & -0.010 & 0.048 & 0.944 \\
 & 1000 & 0.015 & 0.202 & 0.954 & -0.002 & 0.035 & 0.953 & 0.003 & 0.028 & 0.947 & -0.007 & 0.033 & 0.938 \\
\cmidrule{1-14}
SN, $\gamma=2.5$ & 100 & 0.007 & 1.986 & 0.549 & -0.005 & 0.125 & 0.560 & 0.040 & 0.092 & 0.524 & -0.079 & 0.139 & 0.520 \\
 & 500 & 0.070 & 0.526 & 0.944 & -0.000 & 0.037 & 0.940 & 0.007 & 0.028 & 0.947 & -0.017 & 0.037 & 0.921 \\
 & 1000 & 0.021 & 0.348 & 0.937 & 0.001 & 0.025 & 0.942 & 0.003 & 0.020 & 0.946 & -0.008 & 0.025 & 0.934 \\
\cmidrule{1-14}
ST, $\gamma=1.5$ & 100 & 0.016 & 0.831 & 0.866 & -0.015 & 0.129 & 0.855 & 0.030 & 0.088 & 0.849 & -0.052 & 0.128 & 0.853 \\
 & 500 & 0.020 & 0.281 & 0.951 & -0.001 & 0.043 & 0.940 & 0.005 & 0.033 & 0.935 & -0.008 & 0.041 & 0.940 \\
 & 1000 & 0.010 & 0.202 & 0.935 & -0.001 & 0.029 & 0.956 & 0.003 & 0.024 & 0.937 & -0.005 & 0.027 & 0.940 \\
\cmidrule{1-14}
SL, $\gamma=1.5$ & 100 & 0.128 & 0.711 & 0.853 & -0.011 & 0.126 & 0.862 & 0.022 & 0.087 & 0.862 & -0.037 & 0.117 & 0.854 \\
 & 500 & 0.009 & 0.226 & 0.912 & -0.001 & 0.039 & 0.909 & 0.003 & 0.031 & 0.913 & -0.004 & 0.037 & 0.906 \\
 & 1000 & 0.005 & 0.150 & 0.929 & -0.001 & 0.027 & 0.926 & 0.002 & 0.021 & 0.928 & -0.002 & 0.025 & 0.932 \\
\bottomrule
\end{tabular}}%
\ifdim\wd\mctabbox>\textwidth\resizebox{\textwidth}{!}{\usebox\mctabbox}\else\usebox\mctabbox\fi

\smallskip
\begin{minipage}{0.98\textwidth}\footnotesize
Design with separated seasonal factors: $c=0.20$, $\phi=0.50$, $\theta=0.30$, $\Phi=0.60$, $\Theta=0.40$. Cov.\ is the empirical coverage of the nominal 95\% Wald interval; with 1,000 replications its Monte Carlo standard error is 0.007 at the nominal level.
\end{minipage}
\end{table}

\begin{table}[!htbp]
\centering\small
\caption{Monte Carlo recovery of the skewness, scale, and tail parameters, design with separated seasonal factors.}
\label{tab:mc_shape}
\ifdefined\mctabbox\else\newsavebox\mctabbox\fi
\sbox\mctabbox{%
\begin{tabular}{ll *{13}{r}}
\toprule
& & & \multicolumn{4}{c}{$\gamma$} & \multicolumn{4}{c}{$\sigma$} & \multicolumn{4}{c}{$\nu$}\\
\cmidrule(lr){4-7}\cmidrule(lr){8-11}\cmidrule(lr){12-15}
Scenario & $n$ & Bnd. & Med. & Bias & RMSE & Cov. & Med. & Bias & RMSE & Cov. & Med. & Bias & RMSE & Cov.\\
\midrule
SN, $\gamma=1$ & 100 & 0.007 & 0.995 & -0.009 & 0.230 & 0.895 & 0.938 & -0.081 & 0.155 & 0.925 & -- & -- & -- & -- \\
 & 200 & 0.000 & 0.994 & -0.007 & 0.123 & 0.929 & 0.976 & -0.028 & 0.065 & 0.931 & -- & -- & -- & -- \\
 & 500 & 0.000 & 0.999 & 0.002 & 0.072 & 0.939 & 0.991 & -0.009 & 0.034 & 0.950 & -- & -- & -- & -- \\
 & 1000 & 0.000 & 1.006 & 0.005 & 0.049 & 0.937 & 0.995 & -0.005 & 0.023 & 0.948 & -- & -- & -- & -- \\
\cmidrule{1-15}
SN, $\gamma=1.5$ & 100 & 0.062 & 1.498 & 0.001 & 0.265 & 0.857 & 0.963 & -0.084 & 0.261 & 0.919 & -- & -- & -- & -- \\
 & 200 & 0.001 & 1.499 & 0.005 & 0.144 & 0.928 & 0.989 & -0.031 & 0.138 & 0.928 & -- & -- & -- & -- \\
 & 500 & 0.000 & 1.496 & -0.002 & 0.078 & 0.947 & 0.998 & -0.007 & 0.069 & 0.939 & -- & -- & -- & -- \\
 & 1000 & 0.000 & 1.498 & -0.002 & 0.052 & 0.953 & 1.002 & -0.002 & 0.047 & 0.949 & -- & -- & -- & -- \\
\cmidrule{1-15}
SN, $\gamma=2.5$ & 100 & 0.403 & 2.685 & -0.179 & 0.310 & 0.805 & 0.876 & 0.193 & 0.374 & 0.755 & -- & -- & -- & -- \\
 & 200 & 0.114 & 2.472 & -0.026 & 0.190 & 0.910 & 1.014 & 0.017 & 0.275 & 0.896 & -- & -- & -- & -- \\
 & 500 & 0.004 & 2.445 & -0.010 & 0.119 & 0.907 & 1.036 & 0.010 & 0.179 & 0.900 & -- & -- & -- & -- \\
 & 1000 & 0.000 & 2.486 & -0.002 & 0.076 & 0.945 & 1.011 & 0.001 & 0.115 & 0.933 & -- & -- & -- & -- \\
\cmidrule{1-15}
ST, $\gamma=1.5$ & 100 & 0.096 & 1.504 & 0.012 & 0.207 & 0.899 & 0.962 & -0.070 & 0.244 & 0.938 & 4.34 & -0.107 & 1.420 & 0.965 \\
 & 200 & 0.007 & 1.496 & 0.007 & 0.129 & 0.927 & 0.991 & -0.020 & 0.144 & 0.940 & 4.16 & 0.108 & 0.785 & 0.998 \\
 & 500 & 0.000 & 1.492 & -0.001 & 0.074 & 0.941 & 0.997 & -0.004 & 0.082 & 0.951 & 4.04 & 0.034 & 0.407 & 0.965 \\
 & 1000 & 0.000 & 1.492 & -0.002 & 0.049 & 0.936 & 1.006 & 0.002 & 0.056 & 0.939 & 4.04 & 0.022 & 0.274 & 0.958 \\
\cmidrule{1-15}
SL, $\gamma=1.5$ & 100 & 0.001 & 1.470 & -0.005 & 0.169 & 0.882 & 0.982 & -0.043 & 0.190 & 0.928 & -- & -- & -- & -- \\
 & 200 & 0.000 & 1.501 & 0.003 & 0.104 & 0.889 & 0.988 & -0.018 & 0.114 & 0.923 & -- & -- & -- & -- \\
 & 500 & 0.000 & 1.500 & 0.001 & 0.057 & 0.918 & 0.997 & -0.005 & 0.062 & 0.944 & -- & -- & -- & -- \\
 & 1000 & 0.000 & 1.497 & -0.001 & 0.037 & 0.947 & 0.999 & -0.003 & 0.045 & 0.936 & -- & -- & -- & -- \\
\bottomrule
\end{tabular}}%
\ifdim\wd\mctabbox>\textwidth\resizebox{\textwidth}{!}{\usebox\mctabbox}\else\usebox\mctabbox\fi

\smallskip
\begin{minipage}{0.98\textwidth}\footnotesize
Design with separated seasonal factors; $\sigma=1$ and, for ST, $\nu=4$. Bnd.\ is the proportion of boundary fits (an AR or MA coefficient above 0.98 in absolute value, $\hat\gamma\notin[0.25,4]$, or $\hat\nu>150$). Med.\ is the median estimate over all replications. Bias, RMSE, and coverage of the nominal 95\% Wald interval are computed for $\log\gamma$, $\log\sigma$, and $\log(\nu-2)$ over the non-boundary fits.
\end{minipage}
\end{table}

Tables~\ref{tab:mc_recovery} and~\ref{tab:mc_shape} report the separated design. At $n=1000$ the absolute bias is at most 0.008 for $\phi$, $\theta$, $\Phi$, and $\Theta$ and at most 0.021 for $c$ in the eight scenarios, and the coverage of the Wald intervals lies between 0.926 and 0.964. At $n=500$ it lies between 0.906 and 0.957. At $n=200$ it is between 0.90 and 0.95 for the SN and ST members with $\gamma\le 1.5$, between 0.87 and 0.89 for the SL member, and between 0.78 and 0.84 for the SN member with $\gamma=2.5$. The outer-product standard errors of the SL member are therefore optimistic in short series.

The distribution parameters behave similarly. For $n\ge 200$ the bias of $\log\hat\gamma$ and $\log\hat\sigma$ over non-boundary fits does not exceed 0.031 in absolute value, and at $n=1000$ their coverage lies between 0.933 and 0.959. The median of $\hat\nu$ is 4.01 to 4.04 at $n=1000$, with coverage between 0.958 and 0.964; for $n\le 200$ the intervals for $\nu$ are very wide and overcover. Symmetric and nearly symmetric innovations cause no numerical difficulty.

\begin{figure}[!htbp]
\centering
\includegraphics[width=\textwidth]{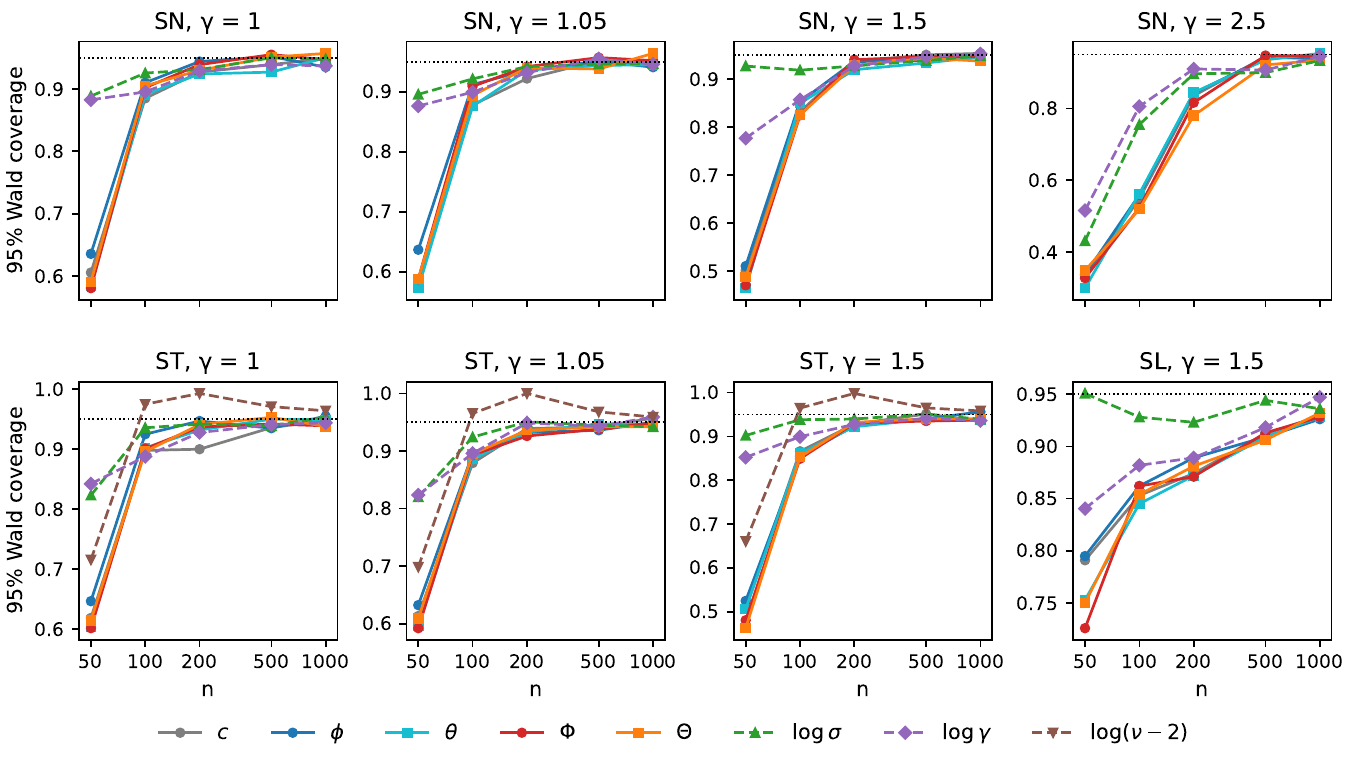}
\caption{Coverage of nominal 95\% Wald intervals in the design with separated seasonal factors. Solid lines: intercept and autoregressive and moving-average coefficients, all valid fits. Dashed lines: $\log\sigma$, $\log\gamma$, and $\log(\nu-2)$, non-boundary fits. The dotted line marks 0.95.}
\label{fig:mc_coverage}
\end{figure}

Short series with strong asymmetry do. With $\gamma=2.5$ the proportion of boundary fits is 65.6\%, 40.3\%, 11.4\%, and 0.4\% at $n=50$, 100, 200, and 500. This pattern is consistent with samples in which few residuals fall below the mode, so that the two-piece likelihood is maximized by a nearly half-normal fit. At $n=50$ boundary fits occur in 13\% to 66\% of the replications of every scenario, and the full seasonal specification should not be fitted to series of that length. Figure~\ref{fig:mc_coverage} summarizes the approach of the coverage to its nominal level.

\subsection{Comparison with Gaussian SARIMA}\label{subsec:mc_gaussian}

\begin{table}[!htbp]
\centering\small
\caption{Comparison with Gaussian SARIMA: tests of $\gamma=1$, information criteria, and relative efficiency.}
\label{tab:mc_tests}
\ifdefined\mctabbox\else\newsavebox\mctabbox\fi
\sbox\mctabbox{%
\begin{tabular}{ll rr rrr rrrr}
\toprule
& & \multicolumn{2}{c}{Rejection of $\gamma=1$} & \multicolumn{3}{c}{Preference for modal} & \multicolumn{4}{c}{Relative efficiency}\\
\cmidrule(lr){3-4}\cmidrule(lr){5-7}\cmidrule(lr){8-11}
Scenario & $n$ & Wald & LR & AIC & BIC & $\overline{\Delta\mathrm{AIC}}$ & $\phi$ & $\theta$ & $\Phi$ & $\Theta$\\
\midrule
SN, $\gamma=1$ & 200 & 0.071 & 0.068 & 0.179 & 0.037 & -0.9 & 0.96 & 0.96 & 0.97 & 0.96 \\
 & 1000 & 0.063 & 0.063 & 0.159 & 0.010 & -0.9 & 1.00 & 1.00 & 1.00 & 1.00 \\
\cmidrule{1-11}
SN, $\gamma=1.05$ & 200 & 0.102 & 0.100 & 0.233 & 0.053 & -0.6 & 0.97 & 0.96 & 0.97 & 0.98 \\
 & 1000 & 0.172 & 0.174 & 0.351 & 0.046 & -0.0 & 1.01 & 1.00 & 0.99 & 0.99 \\
\cmidrule{1-11}
SN, $\gamma=1.5$ & 200 & 0.906 & 0.915 & 0.966 & 0.862 & 10.3 & 1.09 & 1.12 & 1.21 & 1.08 \\
 & 1000 & 1.000 & 1.000 & 1.000 & 1.000 & 61.5 & 1.29 & 1.26 & 1.22 & 1.24 \\
\cmidrule{1-11}
SN, $\gamma=2.5$ & 200 & 0.997 & 1.000 & 1.000 & 1.000 & 35.1 & 1.80 & 1.92 & 1.79 & 1.41 \\
 & 1000 & 1.000 & 1.000 & 1.000 & 1.000 & 203.8 & 2.43 & 2.60 & 2.44 & 2.19 \\
\cmidrule{1-11}
ST, $\gamma=1$ & 200 & 0.073 & -- & 0.938 & 0.773 & 23.2 & 1.28 & 1.31 & 1.27 & 1.33 \\
 & 1000 & 0.056 & -- & 1.000 & 1.000 & 156.8 & 1.41 & 1.43 & 1.45 & 1.40 \\
\cmidrule{1-11}
ST, $\gamma=1.05$ & 200 & 0.079 & -- & 0.944 & 0.755 & 24.9 & 1.27 & 1.34 & 1.30 & 1.33 \\
 & 1000 & 0.222 & -- & 1.000 & 1.000 & 159.7 & 1.39 & 1.36 & 1.44 & 1.38 \\
\cmidrule{1-11}
ST, $\gamma=1.5$ & 200 & 0.959 & -- & 0.998 & 0.975 & 47.4 & 1.57 & 1.55 & 1.63 & 1.55 \\
 & 1000 & 1.000 & -- & 1.000 & 1.000 & 282.9 & 1.77 & 1.90 & 1.82 & 1.88 \\
\cmidrule{1-11}
SL, $\gamma=1.5$ & 200 & 0.993 & -- & 0.999 & 0.998 & 48.5 & 1.80 & 1.72 & 1.72 & 1.89 \\
 & 1000 & 1.000 & -- & 1.000 & 1.000 & 270.8 & 2.10 & 2.23 & 2.42 & 2.20 \\
\bottomrule
\end{tabular}}%
\ifdim\wd\mctabbox>\textwidth\resizebox{\textwidth}{!}{\usebox\mctabbox}\else\usebox\mctabbox\fi

\smallskip
\begin{minipage}{0.98\textwidth}\footnotesize
Wald: proportion of replications with $|\log\hat\gamma|/\mathrm{se}(\log\hat\gamma)>1.96$. LR: proportion with likelihood-ratio statistic above $\chi^2_{1,0.95}$; it is available for the SN member, in which Gaussian SARIMA is nested. AIC and BIC: proportion of replications in which the criterion prefers the modal model to Gaussian SARIMA of the same order; $\Delta\mathrm{AIC}=\mathrm{AIC}_{\rm Gaussian}-\mathrm{AIC}_{\rm Modal}$. Relative efficiency: mean squared error of the Gaussian conditional least-squares estimator divided by that of the modal ML estimator. Design with separated seasonal factors.
\end{minipage}
\end{table}

Table~\ref{tab:mc_tests} compares the modal model with Gaussian SARIMA of the same order, fitted to the same series. Three conclusions follow.

First, the test of symmetry holds its level from about 200 observations. Under SN innovations with $\gamma=1$, which are Gaussian, the Wald test of $\gamma=1$ rejects at the nominal 5\% level in 7.1\% of the replications at $n=200$ and in 6.3\% at $n=1000$; the likelihood-ratio test rejects in 6.8\% and 6.3\%. Under symmetric ST innovations the Wald rates are 7.3\% and 5.6\%. The test is liberal at $n=100$, with rejection rates of 11\%, and unusable at $n=50$, with rates of 30\% to 35\%. Its power against $\gamma=1.5$ is 0.91 at $n=200$ and 1 from $n=500$; against $\gamma=1.05$ it is 0.17 at $n=1000$.

Second, the information criteria behave as they should. Under Gaussian innovations AIC prefers the modal model in 17.9\% of the replications at $n=200$ and in 15.9\% at $n=1000$, close to the 15.7\% implied by a $\chi^2_1$ statistic, and BIC does so in 3.7\% and 1.0\%. Under $\gamma=1.5$ AIC and BIC prefer the modal model in 96.6\% and 86.2\% of the replications at $n=200$ and in all of them from $n=500$. In the ST scenarios the preference reflects the heavy tails as well as the asymmetry: with $\gamma=1$ it is due to the tails alone.

Third, modal ML loses nothing under symmetry and gains under asymmetry. The relative efficiency of the modal estimator of the four coefficients with respect to Gaussian least squares is 0.96 to 0.97 at $n=200$ and 1.00 at $n=1000$ under Gaussian innovations. Under SN innovations it is 1.22 to 1.29 at $n=1000$ for $\gamma=1.5$ and 2.19 to 2.60 for $\gamma=2.5$. Under ST innovations it is 1.40 to 1.45 for $\gamma=1$ and 1.77 to 1.90 for $\gamma=1.5$, and under SL innovations it is 2.10 to 2.42.

\subsection{Nearly cancelling seasonal factors and the starting transient}\label{subsec:mc_transient}

\begin{table}[!htbp]
\centering\small
\caption{Seasonal coefficients and modal level under separated and nearly cancelling seasonal factors.}
\label{tab:mc_seasonal}
\ifdefined\mctabbox\else\newsavebox\mctabbox\fi
\sbox\mctabbox{%
\begin{tabular}{ll *{9}{r}}
\toprule
Scenario & $n$ & Bias $\hat\Phi$ & Bias $\hat\Theta$ & Corr. & Bias $\hat\Phi+\hat\Theta$ & SD $\hat\Phi$ & Cov. $\Phi$ & Bias $\hat\Phi_{\rm G}$ & Med. $\hat c/\hat\alpha(1)$ & Cov. $c$\\
\midrule
\multicolumn{11}{l}{\emph{Nearly cancelling seasonal factors, $\Phi=0.60$, $\Theta=-0.40$}}\\
SN, $\gamma=1$ & 200 & -0.256 & 0.252 & -0.961 & -0.003 & 0.271 & 0.775 & -0.255 & 0.945 & 0.923 \\
 & 1000 & -0.083 & 0.086 & -0.967 & 0.004 & 0.111 & 0.902 & -0.083 & 1.016 & 0.953 \\
\cmidrule{1-11}
SN, $\gamma=1.5$ & 200 & -0.259 & 0.260 & -0.946 & 0.002 & 0.214 & 0.655 & -0.262 & 1.259 & 0.938 \\
 & 1000 & -0.094 & 0.100 & -0.967 & 0.006 & 0.100 & 0.825 & -0.075 & 1.159 & 0.928 \\
\cmidrule{1-11}
SN, $\gamma=2.5$ & 200 & -0.267 & 0.282 & -0.943 & 0.016 & 0.162 & 0.284 & -0.254 & 1.611 & 0.746 \\
 & 1000 & -0.142 & 0.153 & -0.978 & 0.011 & 0.093 & 0.389 & -0.083 & 1.752 & 0.712 \\
\midrule
\multicolumn{11}{l}{\emph{Separated seasonal factors, $\Phi=0.60$, $\Theta=0.40$}}\\
SN, $\gamma=1$ & 200 & 0.007 & -0.015 & -0.578 & -0.008 & 0.071 & 0.939 & 0.007 & 1.090 & 0.927 \\
 & 1000 & 0.002 & -0.005 & -0.569 & -0.003 & 0.031 & 0.947 & 0.002 & 0.906 & 0.949 \\
\cmidrule{1-11}
SN, $\gamma=1.5$ & 200 & 0.012 & -0.024 & -0.577 & -0.012 & 0.061 & 0.941 & 0.007 & 1.064 & 0.937 \\
 & 1000 & 0.003 & -0.007 & -0.588 & -0.004 & 0.028 & 0.947 & 0.002 & 1.012 & 0.954 \\
\cmidrule{1-11}
SN, $\gamma=2.5$ & 200 & 0.018 & -0.043 & -0.530 & -0.024 & 0.050 & 0.816 & 0.008 & 0.977 & 0.835 \\
 & 1000 & 0.003 & -0.008 & -0.555 & -0.005 & 0.020 & 0.946 & 0.001 & 1.142 & 0.937 \\
\bottomrule
\end{tabular}}%
\ifdim\wd\mctabbox>\textwidth\resizebox{\textwidth}{!}{\usebox\mctabbox}\else\usebox\mctabbox\fi

\smallskip
\begin{minipage}{0.98\textwidth}\footnotesize
Corr.\ is the correlation between $\hat\Phi$ and $\hat\Theta$ across replications. $\hat\Phi_{\rm G}$ is the Gaussian conditional least-squares estimate from the same series. The modal level $c/\alpha(1)$, with $\alpha(1)=(1-\phi)(1-\Phi)$, equals 1 in both designs. 1,000 replications per cell.
\end{minipage}
\end{table}

\begin{figure}[!htbp]
\centering
\includegraphics[width=0.85\textwidth]{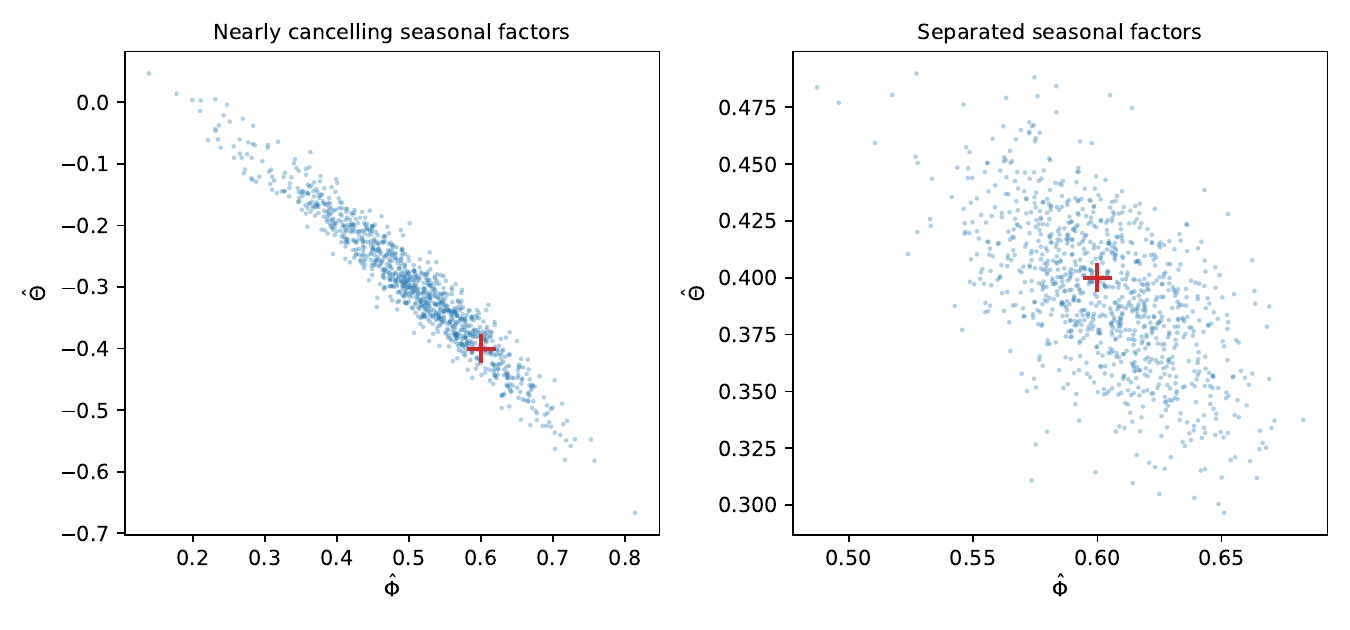}
\caption{Estimates of the seasonal pair in 1,000 replications with SN innovations, $\gamma=1.5$, and $n=1000$. The cross marks the true value.}
\label{fig:mc_ridge}
\end{figure}

The nearly cancelling design behaves differently (Table~\ref{tab:mc_seasonal}). At $n=1000$ the bias of $\hat\Phi$ ranges from $-0.060$ to $-0.142$ over the eight scenarios and the coverage of its interval from 0.907 down to 0.389. The estimates $\hat\Phi$ and $\hat\Theta$ have a correlation between $-0.965$ and $-0.978$ and nearly opposite biases, whereas their sum, which measures the net seasonal effect, has a bias of at most 0.016 for $n\ge 200$ (Figure~\ref{fig:mc_ridge}). The regular coefficients and the distribution parameters are unaffected. The seasonal pair is thus weakly identified along a ridge of the likelihood, as is familiar for nearly cancelling autoregressive and moving-average factors \citep{AnsleyNewbold1980}. For a Gaussian ARMA(1,1) model with coefficients 0.50 and 0.40, \citet{AnsleyNewbold1980} report the same pattern: estimates biased toward the centre of the admissible region and 95\% maximum-likelihood intervals with coverage between 0.34 and 0.43 for series of length 50 and 100. Under symmetric innovations the Gaussian least-squares estimate from the same series has the same bias as the modal estimate ($-0.083$ at $n=1000$). Under $\gamma=2.5$, however, the modal bias is $-0.142$ and the Gaussian one $-0.083$, and the displacement reaches the intercept: the median of the implied modal level $\hat c/\hat\alpha(1)$ is 1.75 instead of 1 and the coverage for $c$ is 0.712.

Proposition~\ref{prop:start} attributes this displacement to the starting values, and Experiment~B confirms it. The same series were fitted under three treatments of the transient: the conditional start, with pre-sample innovations at their mode; a start with pre-sample innovations at $\mathrm{E}(\epsilon_t)$; and the conditional start with the first 60 residuals excluded from the likelihood (Table~\ref{tab:mc_mechanism}). With $\gamma=1.05$ and $n=1000$, excluding the first residuals reduces the bias of $\hat\Phi$ from $-0.076$ to $-0.013$, for the modal and the Gaussian estimators alike. With $\gamma=2.5$ the bias is $-0.139$ under the conditional start, $-0.075$ under the mean start, which is the Gaussian level, and $-0.007$ when the first residuals are excluded; the coverage rises from 0.40 to 0.73 and 0.93, and the bias of $\hat c$ falls from 0.290 to 0.156 and 0.022. The bias under nearly cancelling factors is therefore caused by the starting transient. Starting at the mean brings the bias of $\hat\Phi$ to the Gaussian level but leaves about half of that of $\hat c$, in agreement with the asymmetric variance contribution that a start at $a = m$ does not remove (Section~\ref{sec:conditions}).

\begin{table}[!htbp]
\centering\footnotesize
\caption{Starting transient under nearly cancelling seasonal factors: observed bias and the displacement predicted by Proposition~\ref{prop:start}.}
\label{tab:mc_mechanism}
\ifdefined\mctabbox\else\newsavebox\mctabbox\fi
\sbox\mctabbox{%
\begin{tabular}{lll rrrrrrr}
\toprule
$\gamma$ & $n$ & Start & Bias $\hat\Phi$ & Pred. & Cov. $\Phi$ & Bias $\hat\Phi_{\rm G}$ & Bias $\hat c$ & Pred. & Cov. $c$\\
\midrule
1.05 & 1000 & Mode start & -0.076 & -0.075 & 0.912 & -0.076 & 0.037 & 0.041 & 0.940 \\
 &  & Mean start & -0.075 & -0.074 & 0.912 & -0.076 & 0.036 & 0.040 & 0.940 \\
 &  & Mode start, 60 dropped & -0.013 & -- & 0.916 & -0.013 & 0.005 & -- & 0.934 \\
\cmidrule{2-10}
 & 4000 & Mode start & -0.023 & -0.019 & 0.920 & -0.023 & 0.014 & 0.010 & 0.927 \\
 &  & Mean start & -0.023 & -0.018 & 0.920 & -0.023 & 0.014 & 0.010 & 0.920 \\
 &  & Mode start, 60 dropped & -0.008 & -- & 0.953 & -0.007 & 0.006 & -- & 0.920 \\
\cmidrule{1-10}
2.50 & 1000 & Mode start & -0.139 & -0.366 & 0.400 & -0.083 & 0.290 & 0.894 & 0.754 \\
 &  & Mean start & -0.075 & -0.123 & 0.734 & -0.083 & 0.156 & 0.289 & 0.886 \\
 &  & Mode start, 60 dropped & -0.007 & -- & 0.926 & -0.026 & 0.022 & -- & 0.928 \\
\cmidrule{2-10}
 & 4000 & Mode start & -0.055 & -0.091 & 0.573 & -0.017 & 0.114 & 0.221 & 0.693 \\
 &  & Mean start & -0.022 & -0.031 & 0.873 & -0.017 & 0.042 & 0.071 & 0.907 \\
 &  & Mode start, 60 dropped & -0.002 & -- & 0.973 & -0.000 & -0.001 & -- & 0.960 \\
\bottomrule
\end{tabular}}%
\ifdim\wd\mctabbox>\textwidth\resizebox{\textwidth}{!}{\usebox\mctabbox}\else\usebox\mctabbox\fi

\smallskip
\begin{minipage}{0.98\textwidth}\footnotesize
Nearly cancelling seasonal factors, Skew-Normal innovations. Mode start: pre-sample innovations set to 0. Mean start: pre-sample innovations set to $\mathrm{E}(\epsilon_t)$. 60 dropped: the first 60 residuals are excluded from the likelihood. Pred.: displacement $\{n\,I\}^{-1}b(a)$ of Proposition~\ref{prop:start}, with $b(a)$ and $I$ evaluated by simulation at the true parameter. $\hat\Phi_{\rm G}$: Gaussian conditional least squares. The three fits use the same series; 500 replications for $n=1000$, 150 replications for $n=4000$.
\end{minipage}
\end{table}

Table~\ref{tab:mc_mechanism} also reports the displacement $\{n\,I\}^{-1}b(a)$ implied by Proposition~\ref{prop:start}, with $b(a)$ and $I$ evaluated by simulation at the true parameter. For $\gamma=1.05$ the prediction for $\hat\Phi$ is $-0.075$ at $n=1000$ and $-0.019$ at $n=4000$, against observed biases of $-0.076$ and $-0.023$. For $\gamma=2.5$ the prediction has the correct sign and ordering but overstates the magnitude, and the agreement improves with the sample size: the ratio of observed to predicted bias under the conditional start rises from 0.38 at $n=1000$ to 0.60 at $n=4000$.

The remedies have a cost in short series. At $n=200$ excluding 60 residuals removes a third of the sample, and the coverage for $\Phi$ stays below 0.72. The mean start depends on $\hat\sigma$ and $\hat\gamma$, which are unstable in short and strongly asymmetric series: in the separated design with $\gamma=2.5$ and $n=200$ it lowers the coverage for $c$ from 0.81 to 0.75. In the separated design the three treatments are equivalent from $n=500$ (supplement). In applications, a strongly negative correlation between $\hat\Phi$ and $\hat\Theta$ signals near cancellation. Neither the individual seasonal coefficients nor the intercept should then be interpreted separately, a specification with a single seasonal term should be considered, and in long series the first seasons of residuals may be excluded from the likelihood.

\subsection{Choice of distribution and orders}\label{subsec:mc_selection}

\begin{table}[!htbp]
\centering\small
\caption{Selection of the innovation distribution by information criteria.}
\label{tab:mc_kernel}
\ifdefined\mctabbox\else\newsavebox\mctabbox\fi
\sbox\mctabbox{%
\begin{tabular}{ll rrrr rrrr}
\toprule
& & \multicolumn{4}{c}{Selected by AIC} & \multicolumn{4}{c}{Selected by BIC}\\
\cmidrule(lr){3-6}\cmidrule(lr){7-10}
True innovations & $n$ & G & SN & ST & SL & G & SN & ST & SL\\
\midrule
Gaussian & 200 & 0.796 & 0.176 & 0.006 & 0.022 & 0.956 & 0.036 & 0.000 & 0.008 \\
 & 1000 & 0.814 & 0.158 & 0.028 & 0.000 & 0.990 & 0.010 & 0.000 & 0.000 \\
\cmidrule{1-10}
SN, $\gamma=1.5$ & 200 & 0.040 & 0.908 & 0.038 & 0.014 & 0.174 & 0.804 & 0.000 & 0.022 \\
 & 1000 & 0.000 & 0.924 & 0.076 & 0.000 & 0.000 & 0.996 & 0.004 & 0.000 \\
\cmidrule{1-10}
ST, $\gamma=1.5$, $\nu=4$ & 200 & 0.000 & 0.038 & 0.490 & 0.472 & 0.012 & 0.072 & 0.188 & 0.728 \\
 & 1000 & 0.000 & 0.000 & 0.968 & 0.032 & 0.000 & 0.000 & 0.908 & 0.092 \\
\cmidrule{1-10}
SL, $\gamma=1.5$ & 200 & 0.000 & 0.008 & 0.066 & 0.926 & 0.002 & 0.012 & 0.010 & 0.976 \\
 & 1000 & 0.000 & 0.000 & 0.006 & 0.994 & 0.000 & 0.000 & 0.000 & 1.000 \\
\bottomrule
\end{tabular}}%
\ifdim\wd\mctabbox>\textwidth\resizebox{\textwidth}{!}{\usebox\mctabbox}\else\usebox\mctabbox\fi

\smallskip
\begin{minipage}{0.98\textwidth}\footnotesize
Proportion of replications in which each candidate attains the smallest criterion. G denotes Gaussian SARIMA. All candidates have the true orders and are evaluated on the same residuals; design with separated seasonal factors; 500 replications per cell.
\end{minipage}
\end{table}

In Experiment~C each series from the separated design was fitted with the three members and with Gaussian SARIMA (Table~\ref{tab:mc_kernel}). Under Gaussian innovations BIC selects Gaussian SARIMA in 95.6\% of the replications at $n=200$ and in 99.0\% at $n=1000$, and AIC in about 80\%. SN innovations are identified by AIC in 90.8\% of the replications at $n=200$ and SL innovations in 92.6\%. ST innovations with $\nu=4$ are harder to tell from SL innovations: at $n=200$ AIC selects ST in 49.0\% and SL in 47.2\% of the replications, and the correct rate reaches 96.8\% for AIC and 90.8\% for BIC only at $n=1000$.

The consequences of a wrong choice differ across members (supplement). The coefficient estimates are unbiased under every fitted member. The ST member is also safe for inference: at $n=1000$ the coverage of its intervals for $\phi$, $c$, and $\gamma$ lies between 0.94 and 0.97 whatever the true distribution. The SN member fitted to ST innovations gives coverages of 0.81 for $c$ and 0.62 for $\gamma$. The SL member fitted to Gaussian or SN innovations gives coverages of 0.76 to 0.78 for $\phi$ and of 0.52 to 0.62 for $c$, so its outer-product standard errors are reliable only when the innovations are Laplace-like. When the member is in doubt, the ST member is the prudent choice.

In Experiment~D the SN member was fitted for the sixteen orders with $p,q,P,Q\in\{0,1\}$, all conditioning on the same residuals (supplement). In the separated design AIC selects the true order in 92.5\% of the replications at $n=200$ and in all of them at $n=500$; BIC does so in 67.5\% and 99.5\%. In the nearly cancelling design the true order is selected by AIC in 33.0\% and 57.0\% of the replications and by BIC in 10.5\% and 22.0\%; BIC instead selects a single seasonal term in 68.0\% and 78.0\% of them, most often the order $(1,1,1)\times(1,1,0)_{12}$. Under near cancellation the criteria thus favour the parsimonious specification recommended above.

\subsection{Contamination}\label{subsec:mc_contamination}

\begin{table}[!htbp]
\centering\footnotesize
\caption{Estimation under innovational and additive contamination.}
\label{tab:mc_robust}
\ifdefined\mctabbox\else\newsavebox\mctabbox\fi
\sbox\mctabbox{%
\begin{tabular}{lll rrrrrrrrrr}
\toprule
Contamination & $n$ & Fitted & Bias $\hat\phi$ & Cov. $\phi$ & Bias $\hat\theta$ & Bias $\hat\Phi$ & Cov. $\Phi$ & Med. $\hat c/\hat\alpha(1)$ & Cov. $c$ & Med. $\hat\gamma$ & Cov. $\gamma$ & Med. $\hat\nu$\\
\midrule
None & 1000 & SN & -0.001 & 0.958 & 0.002 & 0.001 & 0.932 & 1.003 & 0.940 & 1.500 & 0.942 & -- \\
 &  & ST & -0.001 & 0.962 & 0.002 & 0.001 & 0.932 & 0.987 & 0.940 & 1.499 & 0.944 & 202.34 \\
\cmidrule{1-13}
Innovational & 1000 & SN & -0.001 & 0.954 & 0.001 & 0.001 & 0.938 & 6.302 & 0.024 & 1.059 & 0.004 & -- \\
 &  & ST & -0.001 & 0.950 & 0.001 & 0.002 & 0.918 & 2.559 & 0.580 & 1.324 & 0.324 & 3.16 \\
\cmidrule{1-13}
Additive & 1000 & SN & 0.260 & 0.312 & -0.901 & 0.088 & 0.472 & 7.130 & 0.332 & 1.033 & 0.000 & -- \\
 &  & ST & 0.122 & 0.262 & -0.505 & 0.078 & 0.368 & 5.474 & 0.106 & 1.140 & 0.000 & 2.35 \\
\bottomrule
\end{tabular}}%
\ifdim\wd\mctabbox>\textwidth\resizebox{\textwidth}{!}{\usebox\mctabbox}\else\usebox\mctabbox\fi

\smallskip
\begin{minipage}{0.98\textwidth}\footnotesize
Base model: separated seasonal factors, SN innovations with $\gamma=1.5$; true modal level 1. Innovational: $N(0,6^2)$ added to 5\% of the innovations. Additive: $N(0,5^2)$ added to 2\% of the observed series $Y_t$. 500 replications per cell.
\end{minipage}
\end{table}

Experiment~E contaminates the separated design with SN innovations and $\gamma=1.5$ in two ways (Table~\ref{tab:mc_robust}). Under innovational contamination, in which an $N(0,6^2)$ perturbation is added to 5\% of the innovations, the innovations remain independent and the coefficients are estimated without bias and with coverage between 0.92 and 0.95 by both members. The distribution parameters and the intercept are not: at $n=1000$ the SN member gives a median $\hat\gamma$ of 1.06 and a coverage of 0.02 for $c$. The ST member, with a median $\hat\nu$ of 3.2, gives a median $\hat\gamma$ of 1.32 and a coverage of 0.58 for $c$; it mitigates the contamination without removing its effect.

Additive outliers are more harmful. When an $N(0,5^2)$ perturbation is added to 2\% of the observed series $Y_t$, each outlier reaches four values of the differenced series. At $n=1000$ the bias of $\hat\theta$ is $-0.90$ for the SN member and for Gaussian least squares and $-0.51$ for the ST member, and the coverage for $\phi$ is 0.31 and 0.26. No member is robust to additive outliers, which should be detected and adjusted before the model is fitted. On uncontaminated SN series the ST fit coincides with the SN fit, with $\hat\nu$ at its upper bound.

\subsection{Forecasting}\label{subsec:mc_forecast}

Experiment~F evaluates forecasts of $w_{n+h}$ for $h\le 12$. In a simulation the conditional density of $w_{n+h}$ given the past is known: by~\eqref{eq:forecast_error} it is the density of $\sum_{j<h}\psi_j\epsilon_{n+h-j}$ centred at the path generated with future innovations at zero, and it is computed by numerical convolution. Forecasts can therefore be scored without the noise of the future observation. We report the relative density, that is, the true conditional density at the forecast divided by its maximum; the RMSE with respect to the true marginal conditional mode; and the true conditional probability content of nominal 95\% prediction intervals. Three forecasts are compared: the recursive modal forecast~\eqref{eq:forecast}, which is the joint mode of the future path; the marginal modal forecast~\eqref{eq:marginal_mode}, with $\delta_h$ estimated from the fitted model; and the forecast of Gaussian SARIMA, which targets the conditional mean.

\begin{table}[!htbp]
\centering\footnotesize
\caption{Accuracy of joint and marginal modal forecasts and of the Gaussian mean forecast.}
\label{tab:mc_forecast}
\ifdefined\mctabbox\else\newsavebox\mctabbox\fi
\sbox\mctabbox{%
\begin{tabular}{lll r rrr rrr rr}
\toprule
& & & & \multicolumn{3}{c}{Relative density} & \multicolumn{3}{c}{RMSE} & \multicolumn{2}{c}{Coverage}\\
\cmidrule(lr){5-7}\cmidrule(lr){8-10}\cmidrule(lr){11-12}
Innovations & $n$ & $h$ & $\delta_h$ & Joint & Marg. & Mean & Joint & Marg. & Mean & Modal & Gaussian\\
\midrule
SN, $\gamma=1.5$ & 1000 & 1 & 0.000 & 0.994 & 0.994 & 0.906 & 0.110 & 0.110 & 0.724 & 0.949 & 0.953 \\
 &  & 2 & 0.920 & 0.832 & 0.997 & 0.972 & 0.907 & 0.115 & 0.394 & 0.949 & 0.952 \\
 &  & 6 & 1.500 & 0.634 & 0.998 & 0.979 & 1.484 & 0.116 & 0.351 & 0.949 & 0.951 \\
 &  & 12 & 1.540 & 0.620 & 0.997 & 0.978 & 1.521 & 0.120 & 0.353 & 0.948 & 0.951 \\
\cmidrule{1-12}
SN, $\gamma=2.5$ & 1000 & 1 & 0.000 & 0.985 & 0.985 & 0.800 & 0.165 & 0.165 & 2.425 & 0.949 & 0.953 \\
 &  & 2 & 3.250 & 0.446 & 0.998 & 0.940 & 3.208 & 0.171 & 1.139 & 0.948 & 0.957 \\
 &  & 6 & 5.170 & 0.103 & 0.998 & 0.953 & 5.112 & 0.195 & 1.050 & 0.949 & 0.955 \\
 &  & 12 & 5.300 & 0.089 & 0.998 & 0.952 & 5.241 & 0.198 & 1.049 & 0.948 & 0.955 \\
\cmidrule{1-12}
ST, $\gamma=1.5$ & 1000 & 1 & 0.000 & 0.991 & 0.991 & 0.829 & 0.117 & 0.117 & 0.913 & 0.949 & 0.954 \\
 &  & 2 & 0.860 & 0.850 & 0.997 & 0.906 & 0.870 & 0.135 & 0.790 & 0.948 & 0.953 \\
 &  & 6 & 1.580 & 0.643 & 0.997 & 0.928 & 1.591 & 0.143 & 0.746 & 0.948 & 0.953 \\
 &  & 12 & 1.620 & 0.627 & 0.997 & 0.927 & 1.630 & 0.148 & 0.756 & 0.948 & 0.953 \\
\bottomrule
\end{tabular}}%
\ifdim\wd\mctabbox>\textwidth\resizebox{\textwidth}{!}{\usebox\mctabbox}\else\usebox\mctabbox\fi

\smallskip
\begin{minipage}{0.98\textwidth}\footnotesize
Joint: recursive modal forecast~\eqref{eq:forecast}. Marg.: marginal modal forecast~\eqref{eq:marginal_mode} with estimated $\delta_h$. Mean: forecast of Gaussian SARIMA. Relative density: true conditional density of $w_{n+h}$ at the forecast divided by its maximum, averaged over replications. RMSE: root mean squared distance to the true marginal conditional mode. Coverage: true conditional probability content of the nominal 95\% prediction interval, averaged over replications. Design with separated seasonal factors.
\end{minipage}
\end{table}

\begin{figure}[!htbp]
\centering
\includegraphics[width=\textwidth]{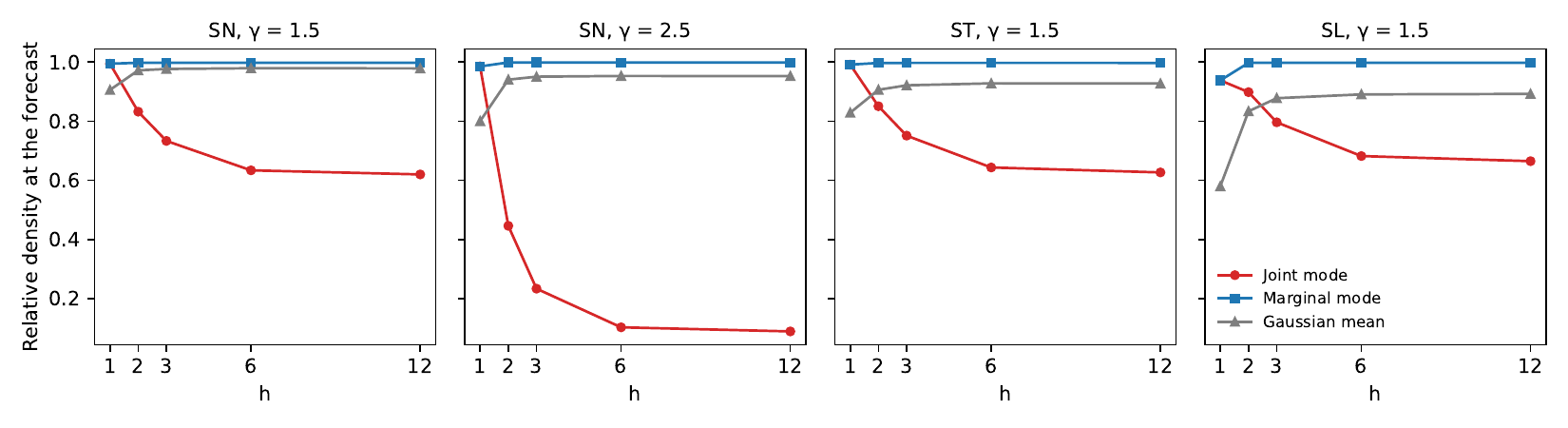}
\caption{Relative density of the true conditional distribution at each point forecast, by horizon; separated design, $n=1000$.}
\label{fig:mc_forecast}
\end{figure}

At one step the two modal forecasts coincide and clearly improve on the mean (Table~\ref{tab:mc_forecast}, Figure~\ref{fig:mc_forecast}). At $n=1000$ their relative density is 0.99 under SN and ST innovations, against 0.91 for the mean under $\gamma=1.5$, 0.80 under $\gamma=2.5$, and 0.83 under ST innovations; under SL innovations it is 0.94 against 0.58.

Beyond one step the recursive forecast is no longer a good forecast of the most probable value. Its relative density falls to 0.62 at $h=12$ under $\gamma=1.5$ and to 0.09 under $\gamma=2.5$, below that of the Gaussian mean, which stays between 0.94 and 0.98 under SN innovations. The marginal modal forecast keeps a relative density of at least 0.997 at every horizon, and its RMSE with respect to the marginal mode is 0.11 to 0.20, against 0.35 to 1.14 for the mean. The correction $\delta_h$ is therefore necessary, not optional, when the forecast is meant to be the most probable value at horizons beyond one.

The modal prediction intervals have a conditional coverage of 0.948 to 0.949 at $n=1000$, with miss probabilities of 0.026 below and 0.025 above. The Gaussian intervals have a coverage of 0.951 to 0.957, but the misses are unbalanced: at most 0.014 below and 0.035 to 0.047 above. At $n=200$ the results are similar except under $\gamma=2.5$, where boundary fits lower the coverage of the modal intervals to between 0.88 and 0.90. The nearly cancelling design leads to the same conclusions (supplement): the weak identification of the seasonal pair does not harm the forecasts.

\subsection{Summary of the simulation evidence}\label{subsec:mc_summary}

With well-separated seasonal factors, conditional ML recovers every parameter and its Wald intervals are reliable from about 200 observations, or 500 under strong asymmetry. The test of symmetry holds its level from the same sample size, the modal estimator is as efficient as Gaussian least squares under symmetry, and it is 1.2 to 2.6 times more efficient under asymmetry or heavy tails. With nearly cancelling seasonal factors the starting transient displaces the seasonal pair and, under asymmetry, the intercept; the net seasonal effect and the forecasts are unaffected. The ST member is the safe choice when the innovation distribution is uncertain, no member is robust to additive outliers, and forecasts of the most probable value beyond one step require the marginal correction.

\section{Application: Canadian Lynx Trappings}\label{sec:application}

We analyse the annual numbers of Canadian lynx trapped in the Mackenzie River district from 1821 to 1934 \citep{EltonNicholson1942, CampbellWalker1977} on the $\log_{10}$ scale ($T = 114$).
The series is stationary (the KPSS-based \texttt{ndiffs} procedure gives $d = 0$), but it is dominated by the well-known ten-year population cycle.
Although the data are annual, this cycle can be represented by a seasonal component of period $s = 10$, which the non-seasonal special case~\eqref{eq:arima_lag} cannot capture beyond its autoregressive terms.
We keep $D = 0$, so that the cycle is modelled as a stationary seasonal autoregression rather than through seasonal differencing, and the non-seasonal models are nested in the seasonal ones.
All models were fitted using the R package \texttt{ModalForecast} (version 0.2.0), which implements the estimation procedure described in Section~\ref{sec:likelihood}. The code accompanying this paper enables reproduction of the results reported here.

For each member of the SKD family, we used the AIC to select the best Modal ARIMA$(p,0,q)$ model with $p \le 4$ and $q \le 2$ and the best Modal SARIMA$(p,0,q)\times(P,0,Q)_{10}$ model subject to the additional constraints $P, Q \le 1$.
Table~\ref{tab:lynx_ic} presents the results, showing that the seasonal model improves on the non-seasonal one for every distribution and under both criteria: the AIC falls by between 14.6 and 28.2 points and the BIC by between 6.4 and 22.8 points.
For the Skewed Student-$t$ member, the degrees of freedom reach the upper bound of the parameter space ($\hat\nu = 202$), so that this fit reduces to the Skew-Normal one and merely adds a parameter.
For comparison, the Gaussian mean-based ARIMA$(2,0,4)$ model chosen by \texttt{auto.arima} has AIC $= -21.27$.

\begin{table}[!htbp]
\centering
\caption{Model comparison for $\log_{10}$ lynx ($T = 114$). $k$ is the number of parameters; standard errors in parentheses.}
\label{tab:lynx_ic}
\small
\begin{tabular}{llrrrrc}
\toprule
Distribution & Orders & $\ell$ & $k$ & AIC & BIC & $\hat\gamma$ (SE) \\
\midrule
\multicolumn{7}{l}{\emph{Non-seasonal models}} \\
Skew-Normal & $(4,0,0)$ & 10.95 & 7 & $-7.90$ & 11.25 & 0.699 (0.107) \\
Skewed Student-$t$ & $(4,0,0)$ & 10.90 & 8 & $-5.81$ & 16.08 & 0.698 (0.106) \\
Skewed Laplace & $(4,0,0)$ & 8.07 & 7 & $-2.14$ & 17.01 & 0.631 (0.081) \\
\midrule
\multicolumn{7}{l}{\emph{Seasonal models}, $s = 10$} \\
Skew-Normal & $(4,0,2)\times(1,0,0)_{10}$ & 21.24 & 10 & $-22.48$ & 4.88 & 0.728 (0.095) \\
Skewed Student-$t$ & $(4,0,2)\times(1,0,0)_{10}$ & 21.28 & 11 & $-20.56$ & 9.54 & 0.728 (0.092) \\
Skewed Laplace & $(3,0,2)\times(1,0,0)_{10}$ & 24.19 & 9 & $-30.39$ & $-5.76$ & 0.888 (0.088) \\
\bottomrule
\end{tabular}
\end{table}

\begin{figure}[!htbp]
\centering
\includegraphics[width=\textwidth]{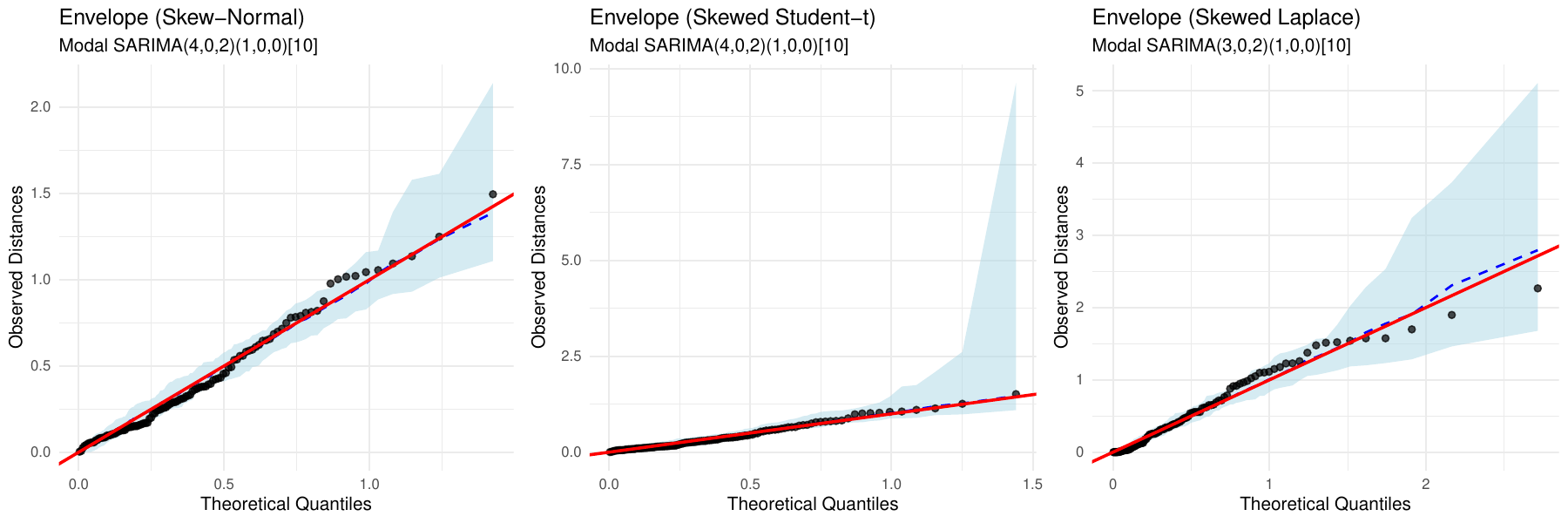}
\caption{Simulation envelopes ($B = 100$, with re-estimation) of the residual distances $\rho_p(\hat\epsilon_t/\hat\sigma)$ for the three Modal SARIMA fits to $\log_{10}$ lynx.}
\label{fig:lynx_envelopes}
\end{figure}

Although the Skewed Laplace SARIMA model has the lowest AIC, three diagnostic checks weigh against its selection.
First, the simulation envelope of its residual distances (Figure~\ref{fig:lynx_envelopes}), built by simulating from the fitted model and re-estimating it in each of $B = 100$ replications, leaves 6 of the 114 distances outside the bands, whereas about $2/(B+1) \approx 2\%$ (i.e., approximately 2.3 points) are expected under a correctly specified model.
Second, its autoregressive polynomial has a pair of complex roots of modulus 1.017, at the edge of the stationarity region of Section~\ref{sec:conditions}.
Third, its skewness is not significant ($z = -1.28$ for $H_0: \gamma = 1$).
In contrast, the Skew-Normal SARIMA model, which ranks second according to the AIC, passes the envelope check (2 points outside), has all its autoregressive roots well outside the unit circle (moduli of 1.17 and 1.36, and 3.47 for the seasonal factor), and exhibits significant skewness. We therefore select it as the final model.

Table~\ref{tab:lynx_est} reports the ML estimates of the selected Skew-Normal Modal SARIMA$(4,0,2)\times(1,0,0)_{10}$ model.
The seasonal autoregressive coefficient is significant ($\hat\Phi_1 = 0.289$, $p = 0.019$), indicating that the ten-year cycle carries information beyond the regular autoregression.
Moreover, the skewness parameter $\hat\gamma = 0.728$ rejects symmetry ($z = -2.86$, $p = 0.004$): the innovations are skewed to the left, so that the conditional mode lies above the conditional mean.
At this sample size the Wald test is somewhat liberal. In the Monte Carlo study of Section~\ref{subsec:mc_gaussian}, with SN innovations, $\gamma = 1$, $n = 100$, and separated seasonal factors, the statistic $(\hat\gamma - 1)/\mathrm{se}(\hat\gamma)$ used here exceeded 1.96 in absolute value in 11\% of the replications and 2.86 in 3\%. The evidence against symmetry therefore remains, although it is weaker than the asymptotic $p$-value suggests.
The randomized quantile residuals show no remaining autocorrelation (Ljung--Box test at lag 20: $p = 0.76$) and are consistent with normality (Shapiro--Wilk test: $p = 0.48$), as expected under a correctly specified model.

\begin{table}[!htbp]
\centering
\caption{Estimates of the Skew-Normal Modal SARIMA$(4,0,2)\times(1,0,0)_{10}$ for $\log_{10}$ lynx. Standard errors from the observed information; delta method for $\sigma$ and $\gamma$.}
\label{tab:lynx_est}
\small
\begin{tabular}{lrrrr}
\toprule
Parameter & Estimate & SE & $z$ & $p$-value \\
\midrule
$c$ & 0.676 & 0.235 & 2.87 & 0.004 \\
$\phi_1$ & 2.142 & 0.123 & 17.39 & $<0.001$ \\
$\phi_2$ & $-2.346$ & 0.215 & $-10.93$ & $<0.001$ \\
$\phi_3$ & 1.312 & 0.217 & 6.04 & $<0.001$ \\
$\phi_4$ & $-0.396$ & 0.126 & $-3.14$ & 0.002 \\
$\theta_1$ & $-1.049$ & 0.100 & $-10.44$ & $<0.001$ \\
$\theta_2$ & 0.892 & 0.055 & 16.09 & $<0.001$ \\
$\Phi_1$ & 0.289 & 0.123 & 2.35 & 0.019 \\
\cmidrule{1-5}
$\sigma$ & 0.182 & 0.019 & & \\
$\gamma$ & 0.728 & 0.095 & & \\
\bottomrule
\end{tabular}
\end{table}

To assess the out-of-sample performance of the selected models, we refitted each model using observations from 1821--1920 and produced forecasts for the 14-year period 1921--1934. Table~\ref{tab:lynx_oos} presents the results.
The seasonal component reduces the mean absolute error (MAE) of the joint modal forecasts for every distribution, from 0.236 to 0.228 for the Skew-Normal and from 0.340 to 0.265 for the Skewed Laplace.
In terms of the root mean squared error (RMSE), which targets the mean rather than the mode, the Gaussian ARIMA model remains the most accurate, and the Skew-Normal SARIMA model is slightly behind its non-seasonal counterpart.
Given the hold-out period of only 14 years, these differences are small; nevertheless, the reduction in absolute error is consistent across distributions.
Note that all 14 held-out values fall inside the 80\% and 95\% prediction intervals of the Skew-Normal SARIMA model.

\begin{table}[!htbp]
\centering
\caption{Hold-out accuracy of joint modal forecasts (and Gaussian mean forecasts) for $\log_{10}$ lynx, 1921--1934.}
\label{tab:lynx_oos}
\small
\begin{tabular}{lrrrr}
\toprule
 & \multicolumn{2}{c}{Non-seasonal} & \multicolumn{2}{c}{Seasonal, $s = 10$} \\
\cmidrule(lr){2-3}\cmidrule(lr){4-5}
Distribution & MAE & RMSE & MAE & RMSE \\
\midrule
Skew-Normal & 0.236 & 0.278 & 0.228 & 0.288 \\
Skewed Student-$t$ & 0.237 & 0.279 & 0.227 & 0.286 \\
Skewed Laplace & 0.340 & 0.378 & 0.265 & 0.308 \\
\cmidrule{1-5}
Gaussian (mean) & 0.225 & 0.258 & 0.247 & 0.277 \\
\bottomrule
\end{tabular}
\end{table}

\begin{figure}[!htbp]
\centering
\includegraphics[width=0.9\textwidth]{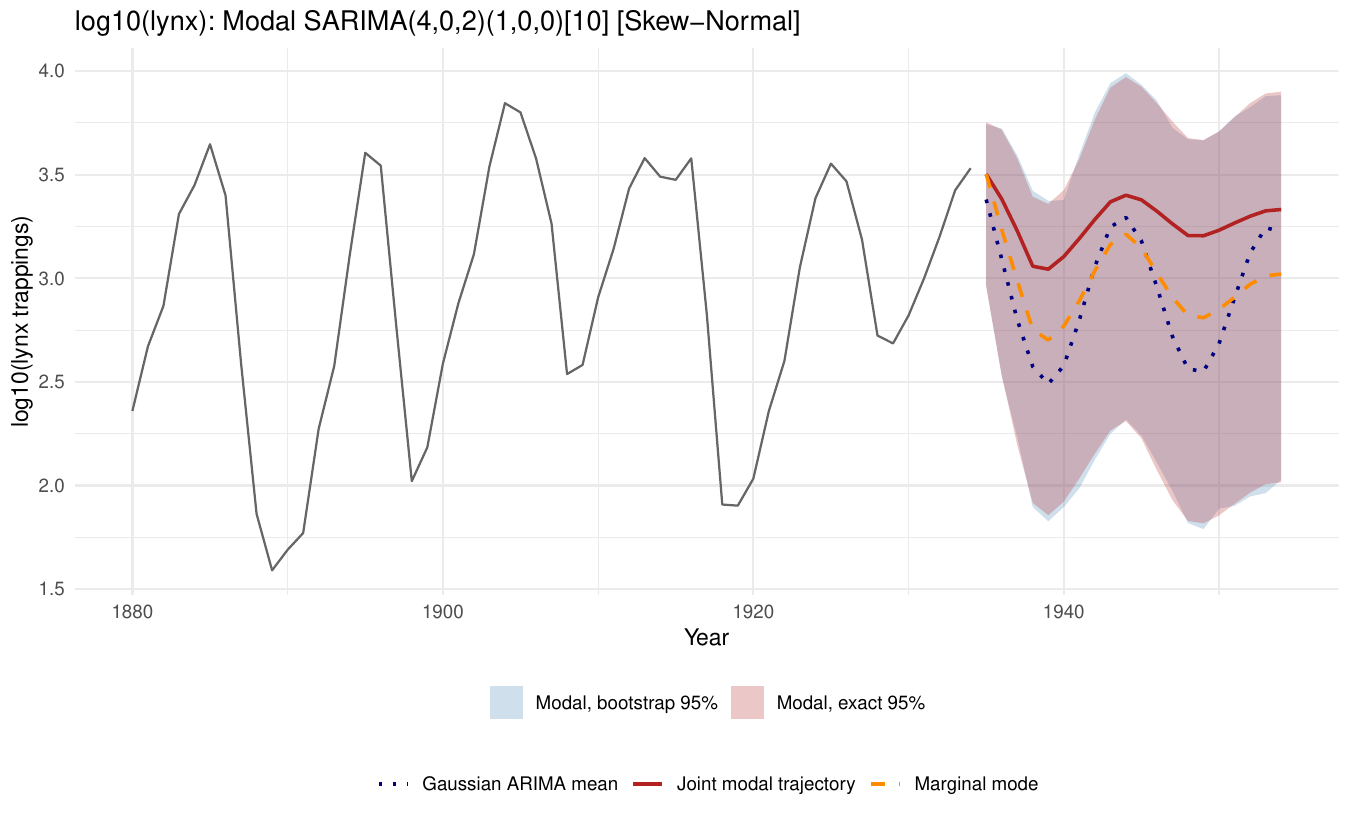}
\caption{Forecasts of $\log_{10}$ lynx for 1935--1954 from the Skew-Normal Modal SARIMA$(4,0,2)\times(1,0,0)_{10}$: joint modal trajectory, marginal mode, Gaussian ARIMA mean, and exact and bootstrap 95\% prediction bands for the modal forecasts.}
\label{fig:lynx_forecast}
\end{figure}

Figure~\ref{fig:lynx_forecast} shows the forecasts for 1935--1954 obtained from the full sample.
The joint modal trajectory, that is, the most probable path, is smoother than the Gaussian mean, whereas the marginal mode of Section~\ref{sec:forecasting} follows the cycle.
For the fitted model the correction $\delta_h$ is negative at every horizon from $h = 2$, so that the marginal mode lies below the joint trajectory: for 1939, the trough of the forecast cycle, the joint trajectory is 3.04, the marginal mode 2.70 and the Gaussian mean 2.49.
Furthermore, the exact prediction intervals, computed from the convolution of the fitted Skew-Normal densities, agree with the 2{,}000-path bootstrap intervals to within 0.05 on the $\log_{10}$ scale at every horizon.

\section{Conclusion}\label{sec:conclusion}

We have proposed the Modal SARIMA$(p,d,q)\times(P,D,Q)_s$ model, in which regular and seasonal autoregressive and moving-average components enter the \emph{conditional-mode} equation through the standard multiplicative polynomial formulation and the innovations follow a skewed distribution with mode zero.
Because the likelihood factorizes across observations and the SKD density depends on the residuals only through the piecewise function $\rho_p(\epsilon_t/\sigma)$, the regular and seasonal structure enters the log-likelihood only through the residual recursion, with lags of orders $p + Ps$ and $q + Qs$.
The estimation framework includes analytical gradients for the Skew-Normal and Student-$t$ members, derivative-free refinement for the nondifferentiable Laplace likelihood, standard errors based on the observed information (or the outer product of scores for the Laplace member), and diagnostic procedures.
Thus, for seasonal time series with asymmetric distributions, the Modal SARIMA model produces forecasts along the \emph{most probable} trajectory rather than forecasts of the conditional mean, while retaining the full interpretive and diagnostic framework of the Box--Jenkins seasonal methodology.
Since the mode is not a linear operator, this joint modal trajectory coincides with the marginal conditional mode only at the one-step horizon, and for longer horizons the marginal modal forecast is obtained by adding the correction $\delta_h$ of~\eqref{eq:marginal_mode}.

The analysis of the Canadian lynx data yields three main findings. First, a seasonal component with a period of ten improves every member of the SKD family relative to its non-seasonal counterpart. Second, re-estimated simulation envelopes distinguish among models when information criteria alone do not. Third, at the trough of the forecast cycle, the joint and marginal modal forecasts differ by approximately one-third of a unit on the $\log_{10}$ scale.
The proposed model, automatic order-selection procedure, and diagnostic tools are implemented in the R package \texttt{ModalForecast} (version 0.2.0), which facilitates both practical applications and reproduction of the results reported in this paper.
The Monte Carlo study supports these results and delimits them. With well-separated seasonal factors, conditional ML and its Wald intervals are reliable from about 200 observations, the test of symmetry holds its level, and the modal estimator is as efficient as Gaussian least squares under symmetry and 1.2 to 2.6 times more efficient under asymmetry or heavy tails. With nearly cancelling seasonal factors the starting transient displaces the seasonal coefficients and, under asymmetry, the intercept, as Proposition~\ref{prop:start} predicts, while the net seasonal effect and the forecasts are unaffected. The Student-$t$ member is the prudent choice when the innovation distribution is uncertain, no member is robust to additive outliers, and beyond one step the marginal correction $\delta_h$ is required for the forecast to remain the most probable value.
Several extensions merit investigation, including applications to macroeconomic and environmental series with strong seasonal components, robust contamination models, and improved small-sample inference for seasonal coefficient estimates.
A detailed investigation of these extensions is beyond the scope of the present paper and remains a promising direction for future research.

\appendix
\renewcommand{\theequation}{A.\arabic{equation}}
\setcounter{equation}{0}
\section{Proofs}\label{app:proofs}

\begin{proof}[Proof of Proposition~\ref{prop:loglik}]
Given the pre-sample values, iterating~\eqref{eq:mode_explicit} maps $(\epsilon_t)_{t \in \mathcal{T}}$, which is independent of them by causality, to $(w_t)_{t \in \mathcal{T}}$ by an affine lower-triangular transformation with unit diagonal.
The conditional density of $(w_t)_{t \in \mathcal{T}}$ is therefore $\prod_{t \in \mathcal{T}} f(\epsilon_t \mid 0, \sigma, \gamma; g)$, with $\epsilon_t$ computed from~\eqref{eq:recursion}.
The seasonal structure affects $\epsilon_t$ through the expanded polynomials $\alpha(B)$ and $\beta(B)$, but not the density evaluated at each $\epsilon_t$.
\end{proof}

\begin{proof}[Proof of Proposition~\ref{prop:stationarity}]
Write $\epsilon_t = m + u_t$, where $\{u_t\}$ is independent and identically distributed with mean zero and finite variance by (A2), and hence white noise.
Since $\beta(B)\,m = \beta(1)\,m$, the model~\eqref{eq:modal_sarima} reads
\begin{equation}\label{eq:centered_arma}
    \alpha(B)\,w_t = c + \beta(1)\,m + \beta(B)\,u_t .
\end{equation}

Suppose first that $\alpha(z) \neq 0$ for $|z| \le 1$.
Then $\alpha(1) \neq 0$, and $x_t = w_t - \kappa$, with $\kappa = \{c + \beta(1)\,m\}/\alpha(1)$, satisfies $\alpha(B)\,x_t = \beta(B)\,u_t$, a zero-mean ARMA$(p + Ps, q + Qs)$ equation driven by white noise.
Because $\alpha(z)$ has no zeros in the closed unit disc, $1/\alpha(z)$ has a power series expansion with absolutely summable coefficients on $|z| \le 1$, so that $x_t = \psi(B)\,u_t$ with $\psi(z) = \beta(z)/\alpha(z)$ is a causal stationary solution, and since $\alpha(z) \neq 0$ on $|z| = 1$ it is the unique stationary solution \citep[Theorem~3.1.3]{BrockwellDavis1991}.
Adding $\kappa$ gives~\eqref{eq:causal}.

Conversely, let $\{w_t\}$ be a causal stationary solution and let $x_t = w_t - \mathrm{E}(w_t)$.
By~\eqref{eq:centered_arma}, $\alpha(B)\,x_t - \beta(B)\,u_t$ equals the constant $c + \beta(1)\,m - \alpha(1)\,\mathrm{E}(w_t)$, and this constant is zero because the left-hand side has mean zero.
Thus $\{x_t\}$ is a causal ARMA process with autoregressive polynomial $\alpha(z)$ and moving-average polynomial $\beta(z)$, which have no common zeros by (A1), and Theorem~3.1.1 of \citet{BrockwellDavis1991} gives $\alpha(z) \neq 0$ for $|z| \le 1$.
The equivalence with the conditions on $\phi$ and $\Phi$ follows from~\eqref{eq:root_conditions}.
\end{proof}

\begin{proof}[Proof of Proposition~\ref{prop:invertibility}]
With $x_t = w_t - \kappa$ and $u_t = \epsilon_t - m$ as in the proof of Proposition~\ref{prop:stationarity}, $\{x_t\}$ is a causal ARMA process with $\alpha(B)\,x_t = \beta(B)\,u_t$.
By Theorem~3.1.2 of \citet{BrockwellDavis1991} and (A1), $u_t = \sum_{j \ge 0} \pi_j\,x_{t-j}$ with $\sum_j |\pi_j| < \infty$ if and only if $\beta(z) \neq 0$ for $|z| \le 1$, and then $\pi(z) = \alpha(z)/\beta(z)$.
Substituting $x_{t-j} = w_{t-j} - \kappa$ and $u_t = \epsilon_t - m$ gives~\eqref{eq:invertible} with $\pi^{*} = m - \kappa\sum_j \pi_j = m - \pi(1)\,\kappa$, which equals $-c/\beta(1)$ by the expression for $\kappa$ in Proposition~\ref{prop:stationarity}; this proves the equivalence and~(a).
The equivalence with the conditions on $\theta$ and $\Theta$ follows from~\eqref{eq:root_conditions} applied to $\beta(z)$.

For (b), each $\epsilon_{t-k}$ with $k \ge 1$ is, by~\eqref{eq:invertible}, a function of $w_{t-k}, w_{t-k-1}, \ldots$, and $\mu_t$ in~\eqref{eq:mu_t_sarima} depends only on these residuals and on $w_{t-1}, \ldots, w_{t-p-Ps}$.
Since $\epsilon_t$ is independent of $\mathcal{F}_{t-1}$ by causality, $w_t = \mu_t + \epsilon_t$ has conditional density $f(w_t - \mu_t \mid 0, \sigma, \gamma; g)$, whose mode is $\mu_t$.

For (c), the computed and the true residuals satisfy the same recursion~\eqref{eq:recursion} once the observed values of $w$ are used, so that after the first $p + Ps$ steps their difference solves the homogeneous equation $\beta(B)\,d_t = 0$.
Its solutions are linear combinations of $t^{\ell} z_i^{-t}$, where $z_i$ are the roots of $\beta(z)$ and $\ell$ is smaller than the multiplicity of $z_i$.
Since $|z_i| \ge r > 1$, each term is bounded by a constant times $\rho^{t}$ for any $\rho \in (1/r, 1)$.
\end{proof}

\begin{proof}[Proof of Proposition~\ref{prop:start}]
For $t \le t_0$ the computed residual is $a$, so $d_t = a - \epsilon_t$ and $\mathrm{E}(d_t) = a - m$.
For $t > t_0$ the computed and the true residuals satisfy the same recursion, so $\beta(B)\,d_t = 0$, and taking expectations gives~(a).
Moreover, $d_t$ is a linear combination of the $q + Qs$ values $a - \epsilon_{t_0}, \ldots, a - \epsilon_{t_0-q-Qs+1}$ with coefficients bounded by $C\rho^{\,t-t_0}$, so $\mathrm{E}(d_t^2) \le C\rho^{\,2(t-t_0)}$ by (A2).

Let $\vartheta_i$ be an element of $(c, \phi_1, \ldots, \Theta_Q)$, and write $z_t = \partial\epsilon_t(\bm{\Theta})/\partial\vartheta_i$ and $\hat z_t = \partial\hat\epsilon_t(\bm{\Theta};a)/\partial\vartheta_i$ at $\bm{\Theta}_0$.
The sequence $z_t$ is a linear filter of $\{w_t\}$ or $\{\epsilon_t\}$ with absolutely summable weights and has finite variance.
The difference $\hat z_t - z_t$ equals $-z_t$ for $t \le t_0$ and solves $\beta(B)(\hat z_t - z_t) = -\{\partial\beta(B)/\partial\vartheta_i\}\,d_t$ for $t > t_0$, so $\mathrm{E}\{(\hat z_t - z_t)^2\} \le C\rho^{\,2(t-t_0)}$, with a slightly larger $\rho < 1$ if necessary.
The $i$-th element of $\Delta_n(a)$ is $\sum_{t \in \mathcal{T}} \{s(\epsilon_t + d_t)\,\hat z_t - s(\epsilon_t)\,z_t\}$, and
\[
    |s(\epsilon_t + d_t)\,\hat z_t - s(\epsilon_t)\,z_t| \le L\,|d_t|\,|\hat z_t| + |s(\epsilon_t)|\,|\hat z_t - z_t|.
\]
Since $d_t$ and $\hat z_t - z_t$ depend only on innovations up to time $t_0$, they are independent of $\epsilon_t$ for $t > t_0$, and the Cauchy--Schwarz inequality bounds the expectation of the right-hand side by $C\rho^{\,t-t_0}$.
For the distribution parameters the corresponding summand is $\bm{\lambda}(\epsilon_t + d_t) - \bm{\lambda}(\epsilon_t)$, whose norm has expectation at most $L\,\{1 + \mathrm{E}|\epsilon_t|\}\,\mathrm{E}|d_t| + L\,\mathrm{E}(d_t^2) \le C\rho^{\,t-t_0}$.
The summands of $\Delta_n(a)$ are therefore absolutely summable in $L^1$, which gives the almost sure and $L^1$ convergence and the finiteness of $b(a)$, with $\|\mathrm{E}\{\Delta_n(a)\} - b(a)\| \le C\rho^{\,n}$.
Finally, $\mathrm{E}\{\nabla\ell_n(\bm{\Theta}_0)\} = 0$, because $\mathrm{E}\{s(\epsilon_t)\} = 0$ and $\mathrm{E}\{\bm{\lambda}(\epsilon_t)\} = \bm{0}$ and $z_t$ is $\mathcal{F}_{t-1}$-measurable.
\end{proof}

\paragraph{Second-order terms of $b(a)$.}
Because $\epsilon_t$ is independent of $d_t$ and $\hat z_t$ for $t > t_0$, the elements of $b(a)$ for the coefficients are $\sum_{t > t_0}\mathrm{E}\{S(d_t)\,\hat z_t\}$, where, by the information identity,
\[
    S(d) = \mathrm{E}\{s(\epsilon_t + d)\} = -\mathrm{E}\{s(\epsilon_t)^2\}\,d + \tfrac12\,S''(0)\,d^2 + O(d^3),
\]
with $S''(0) = \mathrm{E}\{s''(\epsilon_t)\}$ taken in the distributional sense.
For the SN member $s$ is linear on each side of zero, with slopes $-4p^2/\sigma^2$ and $-4(1-p)^2/\sigma^2$, so that
\[
    S''(0) = f(0 \mid 0, \sigma, \gamma)\,\{s'(0+) - s'(0-)\} = \frac{16\,p\,(1-p)\,(1-2p)}{\sigma^3\sqrt{2\pi}},
\]
which vanishes only when $\gamma = 1$.
Since $\mathrm{E}(d_t^2) = (a - m)^2 H_t^2 + \mathrm{var}(d_t)$ by Proposition~\ref{prop:start}(a), the variance of the pre-sample innovations contributes to $b(a)$ for every $a$, and through $S''(0)$ this contribution depends on the asymmetry; a start at $a = m$ removes the terms in $a - m$ but not this contribution.

\begin{proof}[Proof of Proposition~\ref{prop:joint_mode}]
Given $\mathcal{F}_T$, iterating~\eqref{eq:mode_explicit} defines an affine map from $(\epsilon_{T+1}, \ldots, \epsilon_{T+H})$ to $(w_{T+1}, \ldots, w_{T+H})$ that is lower triangular with unit diagonal, and hence has unit Jacobian.
The conditional joint density of the future path is therefore $\prod_{h=1}^{H} f(\epsilon_{T+h} \mid 0, \sigma, \gamma; g)$, expressed in terms of the path.
Each factor is maximized at $\epsilon_{T+h} = 0$, which corresponds to the path generated by~\eqref{eq:forecast}.
\end{proof}

\begin{proof}[Proof of Proposition~\ref{prop:delta}]
For~(a), the SN and SL densities are log-concave, because $\rho_p^2$ and $\rho_p$ are convex, and so are the densities of the terms $\psi_j\epsilon_{T+h-j}$.
By the theorem of \citet{Ibragimov1956}, a log-concave distribution is strongly unimodal, that is, its convolution with any unimodal distribution is unimodal, so that the sum is unimodal by induction on the number of terms.
The restrictions of each of these densities to $(-\infty, 0)$ and to $(0, \infty)$ extend to entire functions, Gaussian or exponential, and convolution preserves this property, because the convolution integral splits at $0$ and at $x$ into integrals of entire functions over intervals with endpoints in $\{-\infty, 0, x, \infty\}$.
The density of the sum is therefore real-analytic on each side of zero and, being positive and integrable, is not constant on any interval, so that its mode is unique.

For~(b), $\delta_1 = 0$ because the mode of $\epsilon_{T+1}$ is zero.
When $\gamma = 1$, each term $\psi_j\epsilon_{T+h-j}$ has a density that is symmetric and unimodal about zero, and the density $f$ of $\epsilon_{T+h}$, for which $\psi_0 = 1$, is strictly decreasing on $(0, \infty)$.
By \citet{Wintner1938}, the sum of the remaining terms is symmetric and unimodal about zero, and hence, by Khintchine's representation, a mixture of uniform distributions on intervals $(-u, u)$.
For $x > 0$ and $u > 0$, the density of $\epsilon_{T+h}$ plus a uniform variable on $(-u, u)$ has derivative $\{f(x + u) - f(x - u)\}/(2u) < 0$, because $|x + u| > |x - u|$; mixing over $u$ shows that the density of the sum is strictly decreasing on $(0, \infty)$, and by symmetry zero is its unique mode.

For~(c), take $\sigma = 1$ without loss of generality, and write $C = 4p(1-p)$ and $\psi_1 = -k$ with $k > 0$.
The density of $\epsilon_{T+2} + \psi_1\epsilon_{T+1}$ is $\int f(x + ky)\,f(y)\,dy$, whose derivative at zero is $\int f'(ky)\,f(y)\,dy$.
For $y > 0$ both $ky$ and $y$ are positive, $f'(ky) = 2pC\,g'(2pky)$, and the substitution $u = 2py$ turns this part of the integral into $C^2\int_0^\infty g'(ku)\,g(u)\,du$.
For $y < 0$ both are negative, $f'(ky) = -2(1-p)\,C\,g'(2(1-p)k|y|)$, and the substitution $u = 2(1-p)|y|$ turns that part into $-C^2\int_0^\infty g'(ku)\,g(u)\,du$.
The derivative at zero is therefore zero, and since the density is log-concave for the SN and SL members, zero is a mode, which is unique by~(a).

For~(d), let $\epsilon_0, \epsilon_1, \ldots$ be independent copies of $\epsilon_t$.
The sum has the distribution of $\epsilon_0 + Y_h$, with $Y_h = \sum_{j=1}^{h-1}\psi_j\epsilon_j$, and hence the density $\mathrm{E}\{f(x - Y_h)\}$.
For the three members $f$ is bounded and Lipschitz with some constant $L_f$, and $Y_h \to Y_\infty = \sum_{j \ge 1}\psi_j\epsilon_j$ in $L^1$, so that
\[
    \sup_x \bigl|\mathrm{E}\{f(x - Y_h)\} - \mathrm{E}\{f(x - Y_\infty)\}\bigr| \le L_f\,\mathrm{E}|\epsilon_t|\sum_{j \ge h}|\psi_j| \to 0 .
\]
The limit density is continuous and tends to zero at $\pm\infty$.
If its mode $\delta_\infty$ is unique, its supremum over $|x - \delta_\infty| \ge \eta$ is smaller than its maximum for every $\eta > 0$, and the uniform convergence implies that every maximizer of the density of the sum lies within $\eta$ of $\delta_\infty$ for $h$ large enough.
\end{proof}


\end{document}